\documentclass[11pt,graphicx,amsmath]{article}
\usepackage{amsmath}
\usepackage{graphicx}
\usepackage{bm}
\usepackage[dvips]{color}
\usepackage{amssymb}
\usepackage{amsfonts}
\usepackage{comment}
\usepackage{cite}
\usepackage{todonotes}
\numberwithin{equation}{section}

\def\bl#1\el{\begin{align}#1\end{align}}

\def\l{\left}
\def\r{\right}

\title{Adiabatic regularization of power spectrum and stress tensor of relic gravitational wave without low-frequency distortion}
\author{\small
      \,  Yang  Zhang\thanks{yzh@ustc.edu.cn} ,
      \, Bo Wang \thanks{ymwangbo@mail.ustc.edu.cn}
        \\
 \small   Department of  Astronomy,
 Key Laboratory for Researches in Galaxies and Cosmology,
  \\
 \small    University of Science and Technology of China,
 \\
 \small
  No.96, JinZhai Road, Hefei, Anhui, 230026,  China \\
 }

 \date{}

\def\be{\begin{equation}}
\def\ee{\end{equation}}
\def\ba{\begin{eqnarray}}
\def\ea{\end{eqnarray}}
\def\nn{\nonumber}

\def\bl#1\el{\begin{align}#1\end{align}}
\def\l{\left}
\def\r{\right}

\sf
\large

\begin{document}

\maketitle

\begin{abstract}
\large

Adiabatic regularization is a  method
to remove UV divergences in quantum fields in curved spacetime.
For relic gravitational wave generated during inflation,
regularization on all $k$-modes of the power spectrum to 2nd adiabatic order,
and of the energy density and pressure to 4th order, respectively,
causes low-frequency distortions.
To avoid these, we regularize
only the short modes  inside the horizon during inflation
(corresponding to the present frequencies  $f \gtrsim  10^{9}$Hz),
and keep the long modes  intact.
Doing this does not violate the energy conservation
since the $k$-modes of RGW are independent of each other  during inflation.
The resulting  spectra  are   UV convergent
and simultaneously free of low-frequency distortion,
and these properties remain in the present spectra after evolution,
in contrast to regularization at the present time
which has some distortion or irregularities.
The  spectra generally exhibit quick oscillations in frequency domain,
even if the initial spectra during inflation have no oscillations.
This pattern is due to the interference between
the positive and negative frequency modes developed during  cosmic expansion,
and may be probed by future RGW detections.

\end{abstract}


      Quantum fields  in curved spacetime, Gravitational waves, inflationary universe,

\large

\section{Introduction}

Quantum fields
in Minkowski spacetime have ultra-violet (UV) divergences,
such as the zero-point energy
of an infinite number of $k$-modes of the vacuum fluctuations.
This  UV divergence
is usually removed by the procedure of normal ordering,
which amounts to drop   the vacuum fluctuations as a whole.
This practice is justified   in flat spacetime
since  gravitational effects
of the vacuum energy are not considered.
In  Robertson-Walker  spacetime  quantum fields
also contain the UV divergence of  vacuum fluctuations,
and one has to treat it with care,
because the finite part of the vacuum fluctuations has
important physical effects in curved spacetime \cite{FeynmanHibbs1965}.
For instance,
the vacuum fluctuations of  metric perturbations
during inflation are the origin of cosmological perturbations,
and induce  CMB anisotropies and polarization.
While scalar metric perturbation
serves as the seed for large scale structure formation,
the tensor perturbation forms
a stochastic background of relic gravitational wave (RGW)
\cite{Grishchuk1975A,Grishchuk1975B,FordParker1977GW,Starobinsky1979,
Allen1988,AllenRomaro1999,Giovannini1999,Zhang06A,Zhang06B}.
It has a very broad power spectrum
covering the bands of almost all the current GW detectors.
In particular, its high frequency part ($f\gtrsim  10^{9}$Hz)
is the target of
the high-frequency Gaussian beam detectors \cite{Li2003,TongZhangGaussian}.
RGW  during inflation is a quantum field,
the   power spectrum defined in the vacuum state has UV divergences,
which must be removed.
The adiabatic regularization
\cite{ParkerFulling1974,FullingParkerHu1974,Bunch1980,
AndersonParker1987,BirrellDavies,ParkerToms}
is  useful in removing UV divergences
in the $k$-modes of quantum fields.
Refs.\cite{ParkerToms,Agullo2008}
applied to a scalar field during inflation,
found that the power spectrum is changed at lower frequency.

Ref. \cite{Durrer2009}  suggested that,
for the far low frequency at the end of inflation,
no adiabatic subtraction should be performed.
Ref. \cite{Marozzi2011} argued that the adiabatic regularization
   is not valid for the modes after the horizon exit.
Ref.\cite{UrakawaaStarobinsky2009}  showed that  the counter term  of regularization
becomes negligibly small after the Hubble radius crossing,
so that the regularized power spectrum tends to the unregularized one.
Refs.\cite{Alinea2018,Alinea2015,Alinea2016} confirmed this
by some inflation models.
Refs.\cite{Markkanen2018,MarkkanenTranberg2013} pointed out that
a prescription of adiabatic regularization is not unique
from perspective of  renormalization,
because  the infinities to be absorbed into the bare constants
can always carry along a finite term,
and each different finite term will correspond to
a different scheme of adiabatic regularization.
In our previous work  \cite{WangZhangChen2016} (as Paper I),
we studied adiabatic regularization on RGW,
and examined three prescriptions
of regularization of power spectrum  as well as the stress tensor,
but the low-frequency distortions  were not addressed in details.

The infrared (IR) band   ($10^{-20} \sim 10^{-15}$Hz)
of the power spectrum is related to the spectra $C^{XX}_l$ at $l=(2\sim 3000)$
of CMB anisotropies and polarization
\cite{ZhaoZhang2006,XiaZhang20089A,XiaZhang20089B,CaiZhang2012},
any distortion in this  band will affect the predicted $C^{XX}_l$.
Moreover,   as we shall see, the IR convergence
of spectra  can be even altered  by the all-$k$ regularization.
Therefore,  it is would be desired that
the low frequency portion remains intact
under the adiabatic regularization,
since its original aim is to remove only UV divergences.
So the scheme to carry out regularization has to be carefully constructed.

In this paper,  we shall present a detailed  study of
the low-frequency distortions brought about by   all-$k$ regularization
upon  the power spectrum, and the stress tensor of RGW.
We shall trace the origin of   distortions,
and make a distinction that
only the inside-horizon modes are responsible for UV divergences,
whereas the  outside-horizon modes contain no UV divergences.
So we shall naturally regularize only these high-frequency modes,
and hold the low frequency modes intact.
This inside-horizon scheme of  regularization is legitimate
because, at the level of the linearized Einstein equation,
the $k$-modes of RGW  are independent of each other.
The resulting regularized spectra
are  UV convergent and free of low-frequency distortions as well.
With these as the initial condition,
the evolution will yield the present spectra which are well-behaved.
We shall also examine other possible schemes of regularization
performed at the present stage,
and compare them.
We shall analyze the structure of RGW as a quantum field at present stage
as a result of  evolution from inflation,
particularly demonstrate the oscillatory pattern  due to
interference between the positive and negative frequency modes.
The paper is organized as follows.

In Sect. 2, we introduce the analytical solution of RGW,
   and analyze the UV and IR asymptotic behaviors of
   the power  spectrum,
   the  spectral energy density  and pressure.

In Sect. 3, we examine the all-$k$ adiabatic regularization
     of the spectra during inflation,
    and   demonstrate the low-frequency distortions
    and the occurrence of IR divergences.

In Sect. 4, we remedy this by the scheme of inside-horizon regularization
    during inflation.

In Sect. 5, we let the initial regularized  spectra of Sect.4
    evolve into the  present spectra.

In Sect. 6, we analyze the structure of RGW at the present stage,
   demonstrate the oscillatory pattern in the spectra
   due to the interference.

In Sect. 7, we examine possible regularization at the present time,
   and compare with those in Sect. 5.

Sect. 8 contains the conclusions and discussions.

Appendix A gives   the adiabatic counter terms,
Appendix B gives   the asymptotic expressions of modes at high frequency,
Appendix C lists   the analytical RGW solution
     from inflation up to the present accelerating stage.

We use the unit with $c=\hbar= 1$ in this paper.

\section{ Power spectrum, energy density, pressure of  RGW}

The metric of a flat Robertson-Walker spacetime is written as
\be \label{metric}
ds^2=a^2(\tau)[d\tau^2-(\delta_{ij}+h_{ij})dx^idx^j],
\ee
in synchronous gauge,
where $\tau$  is the conformal time,
and  $h_{ij}$ is  traceless and transverse RGW.
To linear order, the wave equation is $\Box h_{ij}=0$.
As a quantum field,  it is written as
\be\label{Fourier}
h_{ij}  ({\bf x},\tau)=\int\frac{d^3k}{(2\pi)^{3/2}}
    \sum_{s={+,\times}} {\mathop \epsilon  \limits^s}_{ij}(k)
    \left[ a^s_{\bf k}   h^s_k(\tau)e^{i\bf{k}\cdot\bf{x}}
    +a^{s\dagger}_{\bf k}h^{s*}_k(\tau)e^{-i\bf{k}\cdot\bf{x}}\right]       ,
\ee
where $k= |{\bf k}|$ is the comoving wavenumber,
two polarization tensors satisfy
\be\label{polariz}
{\mathop \epsilon  \limits^s}_{ij}(k) \delta_{ij}=0,\,\,\,\,
{\mathop \epsilon  \limits^s}_{ij}(k)  k^i=0, \,\,\,\,
{\mathop \epsilon  \limits^s} \, ^{ij}(k)
{\mathop \epsilon  \limits^{s'}}_{ij}(k) =\delta_{ss'},
\ee
and $a^s_{\bf k}$ and $a^{s\dagger}_{\bf k}$ are
the annihilation and creation operators of graviton
satisfying  the   canonical commutation relation
\be \label{aadagger}
 \left[a^s_{\bf k}, a^{r\dagger}_{\bf k'}\right]
    =  \delta_{sr}\delta^3({\bf k-k'}).
\ee
For RGW,
the two  polarization modes $h^+_k$ and $h^{\times}_k$
are assumed to be independent and  statistically  equivalent,
so that the superscript $s=+, \times$ can be dropped.
The $k$-mode   equation is
\be   \label{evolutionh}
h_k^{''}(\tau)+2\frac{a^{'}(\tau)}{a(\tau)}h_k^{'}(\tau)+k^{2}h_k(\tau)=0 .
\ee
The second order RGW  \cite{WangZhang2017A,WangZhang2017B} is not considered here.
We emphasize that,
at the level of first order metric perturbations,
the RGW equation is homogeneous,
and  these $k$-modes of RGW are {\it independent},
as they do not couple to each other,
and no energy exchanges between different $k$-modes.
Consequently, from statistical perspective,
the modes $h_k$ can be described by a Gaussian process,
the mean is zero and the variance is given by its power spectrum
that we shall calculate soon.
Let
\be \label{hkuk}
h_k(\tau)\equiv \frac{A}{ a(\tau)} u_k(\tau),
\ee
where $A  \equiv \sqrt{32\pi G} = \frac{2}{M_{Pl}}$
with $M_{Pl}\equiv 1/\sqrt{8\pi G}$ being the Planck mass,
determined by the quantum normalization   that
there is a zero point energy $\frac{1}{2}\hbar \omega$
in high frequency limit in each $\bf k$-mode and each polarization of RGW.
The  mode $u_k $ satisfies the wave equation
\be  \label{evolution}
u_k''(\tau)+\left[k^2-\frac{a''(\tau)}{a(\tau)} \right]u_k(\tau)=0.
\ee
For each stage of  cosmic  expansion, i.e.,
inflation, reheating, radiation dominant, matter dominant
and the present accelerating,
the scale factor  is taken as  a power-law  form $a(\tau) \propto \tau^b$
where $b$ is  a constant,
and the exact solution of Eq.(\ref{evolution})
is a combination of Hankel functions,
\be\label{solgen}
u_k(\tau ) =\sqrt{\frac{\pi}{2}} \sqrt{\frac{\sigma}{2k}}
      \big[ C_1  H^{(1)}_{b-\frac{1}{2}}(\sigma)
  + C_2 H^{(2)}_{b-\frac{1}{2}} (\sigma)\big],
\ee
where $\sigma \equiv k \tau $,
and $C_1$, $C_2$ are coefficients  determined by continuity
of $u_k $, $u_k' $ at the transition of two consecutive stages.
Thus, the  analytical solution $h_k(\tau)$ is obtained
for the whole course of evolution\cite{Zhang06A,Zhang06B}.
Appendix C gives the coefficients
for these five expanding stages
by connecting the adjoining stages.
For the inflation,
one has
\be \label{inflation}
a(\tau)=l_0|\tau|^{1+\beta},\,\,\,\,-\infty<\tau\leq \tau_1,
\ee
where two constants $l_0$ and $\beta$ are the parameters of  the model,
$\tau_1$ is the ending time of inflation \cite{Grishchuk1975A,Grishchuk1975B,Zhang06A,Zhang06B}.
For various values of the index $\beta  \sim -2$,
the  scale factor in (\ref{inflation})
 describes a class of inflation models.
The expansion index $\beta$ is related to
the slow-roll parameter $\epsilon$ of inflation
via $\beta = \frac{-2+ \epsilon}{1-\epsilon}$.
In  de Sitter inflation,   $\beta=-2$,
$l_0^{-1}= H$
is the inflation expansion  rate.
Eq.(\ref{evolution}) has a  general solution
\ba \label{hankel}
u_k(\tau ) & = & \sqrt{\frac{\pi}{2}}\sqrt{\frac{x}{2k}}
     \big[ A_1  e^{i\pi (\beta +1)/2} H^{(1)}_{ \beta+ \frac{1}{2} } (x)
          +A_2    e^{-i\pi (\beta +1)/2} H^{(2)}_{\beta + \frac{1}{2}} ( x) \big],
          \nn \\
   &&   \,\,\,\, \,\,\,\,\, \, -\infty  <\tau\leq \tau_1,
\ea
where $x \equiv |k \tau|=-k\tau$, the phase $e^{-i\pi (\beta +1)/2}$ is chosen
for simplicity at high frequency.
Each choice of  $k$-dependent coefficients $(A_1 , A_2 )$  defines
a choice of  quantum state  of RGW  during inflation.
The Wronskian  as a conserved quantity
\be \label{wronsk}
u_k (\tau) u'^{\, *}_k(\tau)  -u_k^* (\tau)  u_k' (\tau) =i
\ee
is required for the modes $u_k$ and $u_k^*$ as
the two independent of solutions of the wave equation,
and it is checked by using
$H^{(1)} _k  \frac{d}{dx} H^{(2)}_k  - H^{(2)}_k  \frac{d}{dx} H^{(1)}_k
   = - \frac{4i}{\pi x}$.
The Wronskian   (\ref{wronsk}) holds also in other expansion stages
when the mode and its derivative are connected continuously.
Application of (\ref{wronsk}) to the mode solution (\ref{hankel})
gives a relation $|A_2 |^2 - |A_1|^2=1$.
By
$ H^{(1)}_{\beta+ \frac{1}{2}} =  H^{(2)\, *}_{\beta+ \frac{1}{2}} $,
the solution  (\ref{hankel}) can be  denoted by
\be\label{uv}
   u_k(\tau ) = A_1  v_k^*(\tau) + A_2 v_k(\tau)  ,
\ee
where
\be\label{u}
v_{k }(\tau )\equiv  \sqrt{\frac{\pi}{2}}\sqrt{\frac{x}{2k}}
  e^{-i\pi \frac{(\beta +1)}{2}}  H^{(2)}_{\beta+ \frac{1}{2}} (x).
\ee
In high frequency limit $k\rightarrow \infty$,
it approaches to
\be \label{uinfl}
v_{k }  \rightarrow \frac{e^{-ik\tau}}{\sqrt{2k}}
\Bigg(1-i\frac{\beta(\beta+1)}{2k\tau}
     -\frac{(\beta+2)(\beta+1)\beta(\beta-1)}{8k^2\tau^2}+ {\cal O} (k^{-3})\Bigg),
\ee
where the leading term  $ \frac{e^{-ik\tau}}{\sqrt{2k}}$
 is identified as
the positive-frequency vacuum  mode in   Minkowski spacetime,
and other terms reflect the effects of the expanding RW spacetime.
(See (\ref{ukexpand}) in Appendix B.)
The conjugate  $v^*$ in (\ref{uv}) is associated with the negative-frequency mode.
In general, the solution  (\ref{hankel})
contains  both positive and negative frequency modes,
and the mode $h_k$ is written as
\ba  \label{solAH}
h_k(\tau)
& = & \frac{H\sqrt{\pi}}{M_{Pl}}|\tau|^{-\beta-\frac12 }
   \l[ A_1  e^{i\pi (\beta +1)/2} H^{(1)}_{ \beta+ \frac{1}{2} } (x)
    + A_2   e^{-i\pi (\beta +1)/2} H^{(2)}_{\beta + \frac{1}{2}} ( x) \r], \nn \\
    && ~~~ ~~~ \,\,\, \, {\rm  for}\, \, -\infty<\tau\leq \tau_1 \, .
\ea
One sees that its  amplitude  during inflation
 is mainly  determined by the ratio    $H/M_{Pl}$.

We work in Heisenberg picture,
in which RGW as  a quantum field operator evolves in time,
whereas
Fock space vector of quantum state does not change with time.
One defines the vacuum  state vector $|0 \rangle$ such that
\be\label{ask}
 a^s_{\bf k}  |0 \rangle =0, \,\,
      {\rm for\   all} \  {\bf k}, \ {\rm and }\   s=+,\times \  .
\ee
The power spectrum  of RGW is defined by
\be \label{defspectrum}
\int_0^{\infty}\Delta^2_t(k,\tau)\frac{dk}{k}   \equiv
\langle0|  h^{ij}(\textbf{x},\tau)h_{ij}(\textbf{x},\tau) |0\rangle  ,
\ee
where the vacuum expectation value
\ba\label{vevcorr}
\langle0|  h^{ij}(\textbf{x},\tau)h_{ij}(\textbf{x},\tau) |0\rangle
         =   \frac{1}{(2\pi)^3} \int d^3k \,
         (  |{  h_k ^+}|^2 +  |{  h_k  ^\times }|^2) ,
\ea
is the auto-correlation function of RGW.
One reads off the spectrum
\be \label{spectrum}
\Delta^2_t(k,\tau)= 2\frac{ k^{3}}{2\pi^2}|h_k(\tau)|^2
= \frac{ k^{3}}{\pi^2a^2}\frac{4}{M_{Pl}^2}|u_k(\tau)|^2,
\ee
where the factor 2 is from the two polarizations $+, \times$.
The definition  (\ref{vevcorr})  applies to any  stage of cosmic expansion.
When the choice
\be\label{BDa10}
 A_1 =0, ~~~  A_2=1,
\ee
is taken for inflation,
the mode $u_k=v_k$,
the state defined by (\ref{ask}) is called the  Bunch-Davies  vacuum.
This specifies  an initial condition of RGW during inflation
which is  used in this paper,
and the power spectrum during  inflation becomes
\be\label{BunchDaviesSpectrum}
\Delta^2_t(k,\tau)
  = \frac{ k^{3}}{\pi^2}\frac{4}{a^2M_{Pl}^2}  |v_k(\tau)|^2
  = \frac{k^{2(\beta+2)} }{ \pi l_0^2   M_{Pl}^2}
    x^{-(2\beta+1)}
   \big|  H^{(2)}_{\beta + \frac{1}{2}} ( x)   \big|^2 ,
\ee
which holds for any time $\tau$ during inflation.
For  the  de Sitter inflation   $\beta=-2$,
$v_{k } = \frac{1}{\sqrt{2k}} \left(1 - \frac{i}{x} \right) e^{-i x}$,
(\ref{BunchDaviesSpectrum}) reduces to
\be   \label{psbet2}
\Delta_t^2(k,\tau)
= \frac{2  k^2}{\pi^2  M_{Pl}^2  a^2(\tau)}
   \left( 1+\frac{1}{(k\tau)^2} \right)  .
\ee
In the long wavelength limit $x \ll 1$,
(\ref{BunchDaviesSpectrum})  becomes the  primordial   spectrum of RGW
\be\label{spectr}
\Delta^2_t(k,\tau)
\simeq   a_t^2 \frac{2}{\pi^2} \left(\frac{H}{M_{Pl}}\right)^2 k^{2 \beta+ 4}
\propto  k^{2 \beta+ 4}
\ee
with $a_t  \simeq 1.01$
for $\beta=-2.02$,
which is often written as
    \cite{KosowskyTurner,WMAPSpergel2003,WMAPPeris2003}
\be \label{initial}
\Delta^2_{t} (k)
= \Delta^2_{R}\, r (\frac{k}{k_{0}})^{ n_t +\frac{1}{2}\alpha_t \ln(\frac{k}{k_0})},
\ee
with the pivot wavenumber $k_0=0.002$Mpc$^{-1}$ for WMAP.
In our model there are relations $n_t=2\beta+4$ and $n_t=n_s-1$ \cite{WangZhangChen2016}.
The observed value of the scalar spectral index $n_s=0.962 \sim 0.972$
gives $\beta  \simeq -2.019\sim -2.014$.
For demonstration purpose,
we shall take $\beta =  -2.02 $ and $\beta=-1.98$.
The scalar curvature spectrum
 $\Delta_{R}^{2} \simeq  2.47 \times10^{-9}$
and the tensor-scalar ratio $r< 0.11$
\cite{Komatsu2011,WMAP9Hinshaw2013}.
This leads to an upper limit
$  \frac{2}{\pi^2} (\frac{H}{M_{Pl}})^2 < 2.4 \times 10^{-10}$,
i.e., $H < 3.6\times 10^{14}$GeV.
To be specific,
we take  $H\sim 3 \times 10^{14}$GeV,
corresponding to an inflation energy scale
$\rho^{1/4}\sim 3.5 \times 10^{16} $GeV.

The frequency  at  time $\tau$ is related to $k$
via $f(\tau) =  k/ 2\pi a(\tau)$.
By the normalization of $a(\tau)$ adopted in this paper,
the present frequency is
\be\label{fk}
f  \simeq   1.7\times10^{-19}\,  k ~ {\rm Hz} .
\ee
In particular,
the conformal wavenumber $k$
of horizon-crossing at the end of inflation
is $k \simeq 1/|\tau_1| $,
which corresponds to  $f =  \frac{a(\tau_1)}{a(\tau_H)} \frac{H}{2\pi}$.
Using $ H\sim 3\times 10^{14}$GeV
and $\frac{a(\tau_H)}{a(\tau_1)}\sim 7\times 10^{28}$
gives $ f  \simeq 10^{9}$Hz.
It should be mentioned that
Paper I  \cite{WangZhangChen2016}
adopted  $f\sim 10^{11}$Hz based on a longer reheating model.

If $\Delta^2_t \propto k^{d}$ for $d \ge 0$ at high frequencies,
the auto-correlation (\ref{defspectrum})  will have UV divergence
coming from the upper limit of integration.
During inflation
the squared absolute mode  $|v_k |^2   \propto k^{-1}, k^{-3}$
at high frequency (see (\ref{expusq})),
so that the UV behavior of the spectrum  is the following
\[
\Delta_t^2  \propto k^2, k^0,
\]
where $k^2$ is the quadratic UV divergence coming from
the Minkowski spacetime modes $ \frac{e^{-ik\tau}}{\sqrt{2k}}$
of (\ref{uinfl}),
and $k^0$ is the logarithmic UV divergence which exists in a RW spacetime.
These UV divergences are to removed by adiabatic regularization.

The IR behavior of the spectrum  as $k \rightarrow 0$ is sensitive to
the index $\beta$,  as indicated by Eq.(\ref{spectr}).
For $\beta<-2$ models which are favored by CMB observations,
the auto-correlation
  at the lower limit $k=0$ of integration
is proportional to the following
\be\label{infpow}
\int_0  k^{2\beta+4} \frac{dk}{k}
            = \infty,
\ee
i.e., the auto-correlation is IR power divergent.
For $\beta=-2$ de Sitter inflation,
$\Delta^2_t  \propto k^0$,
the lower limit  gives
$\propto \int_0    \frac{dk}{k} = - \ln k|_0 = \infty $,
the correlation is  logarithmically IR divergent.
For $\beta > -2$,
$\Delta^2_t \rightarrow 0$ at    $k=0$,
 and the  correlation is IR convergent.
Ideally,
an adiabatic regularization procedure
should not alter  the  IR behaviors,
or at least, should not make an IR  convergent spectrum
into  IR  divergent.
Nevertheless, as we shall see,
this can happen for certain inflation models.

We  examine  the UV and IR behaviors  of
the energy density and pressure of RGW.
There are several definitions of energy-momentum
tensor of RGW arising from different considerations
 \cite{BrillHartle1964,Isaacson1968A,Isaacson1968B,Weinberg1972,
FordParker1977GW,Ford1985,Sahin1990,Abramo1997,Giovannini2006,SuZhang}.
For specific,
we consider the  following one
\cite{Weinberg1972,BrillHartle1964,Isaacson1968A,Isaacson1968B,WangZhangChen2016,SuZhang}
\be\label{tmunu1}
t_{\mu \nu }=\frac{1}{32\pi G}\langle 0|h^{ij}_{\ , \, \mu }h_{ij,\, \nu }|0\rangle ,
\ee
which belongs to  the effective  stress tensor,
and is not covariantly conserved  by itself.
The total energy-momentum tensor is covariantly conserved,
including RGW and
other matter components as well \cite{BrillHartle1964,Isaacson1968A,Isaacson1968B,Abramo1997,Giovannini2006,SuZhang}.
The RGW energy density is
\be \label{energyspectr}
\rho_{gw}  =   t^0\, _0=
\frac{1}{32\pi G a^2} \int \frac{d^3 k}{(2\pi)^3}  \, 2 |h_{k}'(\tau)|^2
=\int^{\infty}_0 \, \rho_k(\tau)\frac{d k}{k} ,
\ee
with the spectral energy density
\be  \label{rhok}
\rho_k(\tau) = \frac{k^3}{\pi^2a^2} \left| (\frac{u_{k}}{a} )'\right|^2 ,
\ee
the RGW pressure   is given by
\be \label{prefp}
p_{gw} =  -\frac{1}{3}t^i\, _i= \frac{1}{96 a^2\pi G}
\int \frac{d^3k}{(2\pi)^3}2k^2|h_k|^2
=\int^{\infty}_0 p_k(\tau)\frac{dk}{k} ,
\ee
with  the   spectral pressure
\ba \label{pressFPth}
p_k(\tau) = \frac{k^5}{3\pi^2a^4} |u_k(\tau)|^2 .
\ea
Other definitions of $\rho_k $ and $p_k $ of RGW
also involve linear combinations of
$|u_k |^2$ and $\left| (\frac{u_{k}}{a} )'\right|^2$
as an essential part.

For  de Sitter inflation,
by (\ref{expusq}) (\ref{expdvprim2}),
the spectral energy density and pressure have  the following simple expressions
\be\label{rhob2}
 \rho_k(\tau)  = \frac{k^3}{\pi^2 a^2 } \,  \frac{k}{2 a^2}
       =  \frac{k^4}{2 \pi^2a^4} \,  ,
\ee
\be\label{pbet2}
p_k(\tau) = \frac{k^5}{3\pi^2a^4}  |v_k(\tau)|^2
  =\frac{k^4}{6 \pi^2 a^4}    \left( 1+\frac{1}{(k\tau)^2} \right) .
\ee
Note that $ \rho' _{k} + 3\frac{a'}{a} (\rho_{k}   +  p_{k} )
 = \frac{a'}{a} \frac{1}{ x^2}  \rho_k  \ne 0$
since  the effective $t_{\mu\nu}$ is not covariantly conserved,
as mentioned earlier.
In high frequency limit,   both $\rho_k$ and $p_k$ are  UV divergent,
and in low frequency limit,
$\rho_k \propto H^4 k^4$ and $p_k \propto H^4 k^2$,
both are IR convergent.
For general $\beta \sim -2$ models,
in high frequency limit, by (\ref{expusq}) (\ref{expdvprim2}),
\be
 \rho_k \simeq  \frac{k^4}{2 \pi^2a^4} ,
 ~~~~
p_{k  }   \simeq \frac{k^4}{2\pi^2a^4} .
\ee
In low frequency limit,
\[
|v_k|^2 \simeq    b_t k^{-1} x^{2\beta+2},
~~~  \left| ( \frac{v_k }{a} ) ' \right|^2
  \simeq  c_t l_0^{-2}|\tau|^2 k^{2 \beta+ 5},
\]
where $b_t \equiv \frac{\sin(\pi  \beta )^2\Gamma
\left(-\beta -\frac{1}{2}\right)^2\Gamma
\left(\beta +\frac{3}{2}\right)^2+\pi ^2}{\pi  2^{2\beta +3}
\Gamma\left(\beta +\frac{3}{2}\right)^2}\simeq \frac12 $,
$c_t \equiv  \frac{\sin(\pi  \beta )^2\Gamma
\left(-\beta -\frac{1}{2}\right)^2\Gamma
\left(\beta +\frac{3}{2}\right)^2+\pi ^2}{\pi  2^{2\beta +3}(2\beta +3)^2
\Gamma\left(\beta +\frac{3}{2}\right)^2} \simeq \frac12 $,
so
\be   \label{enersmallk}
 \rho_k
  \simeq    \frac{c_t}{\pi^2  |\tau|^{2\beta }}
      H^{4}  k^{2\beta+8 }
 \propto H^4 k^{2\beta +8}
  ~~~  \text{for fixed $\tau$}   ,
\ee
\be\label{presssmallk}
p_{k}  \simeq   \frac{ b_t }{ 3\pi^2 |\tau|^{2\beta + 2}}
      \,  H^4   k^{2\beta+ 6}
  \propto    H^4  k^{2\beta+6}
    ~~~ \text{\rm for   fixed $\tau$},
\ee
which are IR convergent.
As we shall see,
the all-$k$ adiabatic regularization will bring  $\rho_k$, $p_k$ into IR divergent.
That is our concern.

\section{Adiabatic regularization  during inflation }

\begin{figure}
\centering
\includegraphics[width=0.9\linewidth]{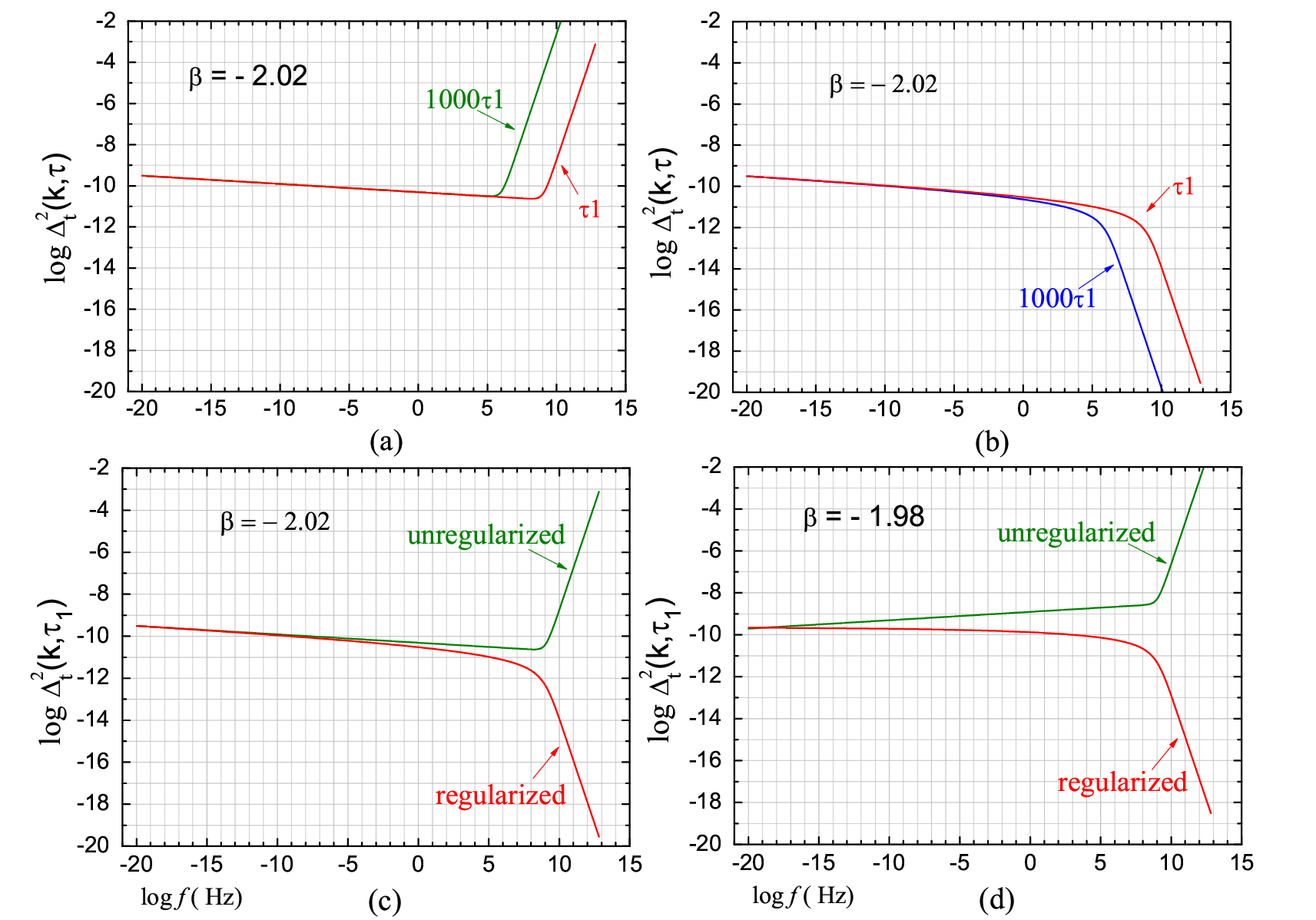}
\caption{
(a) The unregularized power spectrum during inflation.
At an earlier time $\tau=1000\tau_1$
it rises up at $f\sim 10^{6}$Hz,
at the end of inflation $\tau_1$ it rises up at $f_1 \sim 10^{9}$Hz.
(b) The  spectrum regularized
at $1000\tau_1$ and at $\tau_1$ respectively.
The former will evolve into the latter.
(c) (d) The regularized and unregularized spectrum
    at $\tau_1$ for the models  $\beta = -2.02$ and  $\beta = -1.98$.
The horizontal axis is converted into the present frequency $f$ by (\ref{fk}).
      }
     \label{4Rise}
\end{figure}

We now analyze in details the occurrence of
 the low-frequency distortions caused by adiabatic regularization,
which was  not addressed in Paper I.
The UV divergent power spectrum  (\ref{BunchDaviesSpectrum})
 changes with time during inflation.
At a given  instance $\tau$,
the spectrum approximately is flat $\propto  k^{2 \beta+ 4} \sim k^0$
for $k\lesssim  1/|\tau|$
(except at the IR end $k\sim 0$),
and rises up  as $\propto k^2$ for  $k\gtrsim 1/|\tau|$,
i.e., the  horizon-crossing $k|\tau|=1$ is
the point where the spectrum starts to rise up.
By examining the high-frequency expression of $|v_k(\tau)|^2$ in (\ref{expusq}),
we realize  that  the point $k|\tau|=1$ is also
where  the Minkowski spacetime vacuum term $\frac{1}{2k}$
is roughly equal to
the next expansion term $\frac{1}{2k} \frac{\beta(\beta+1)}{2k^2\tau^2}$.
The vacuum term is dominant at $k\gtrsim  1/|\tau|$,
and the expansion terms are dominant for $k\lesssim  1/|\tau|$.
At any time  during inflation,
the part of vacuum modes which lies  inside the horizon
causes the $k^2$ divergence,
and the expansion term causes $k^0$ divergence.
The comoving wavenumber of  horizon-crossing $k=1/|\tau|$
corresponds to a present frequency $f_{\tau}$.
For an earlier time, the rising-up frequency $f_\tau$ is smaller.
Fig.\ref{4Rise} (a) illustrates     that,
at an earlier time $\tau=1000\tau_1$ the spectrum rises up at $f\sim 10^{6}$Hz,
and at the end of inflation $\tau_1$
it rises up at $f\sim 10^{9}$Hz, respectively.
The time $\tau=1000\tau_1$ is when the scale factor is $a \simeq  10^{-3}a(\tau_1)$.

\begin{figure}
\centering
\includegraphics[width=0.9\linewidth]{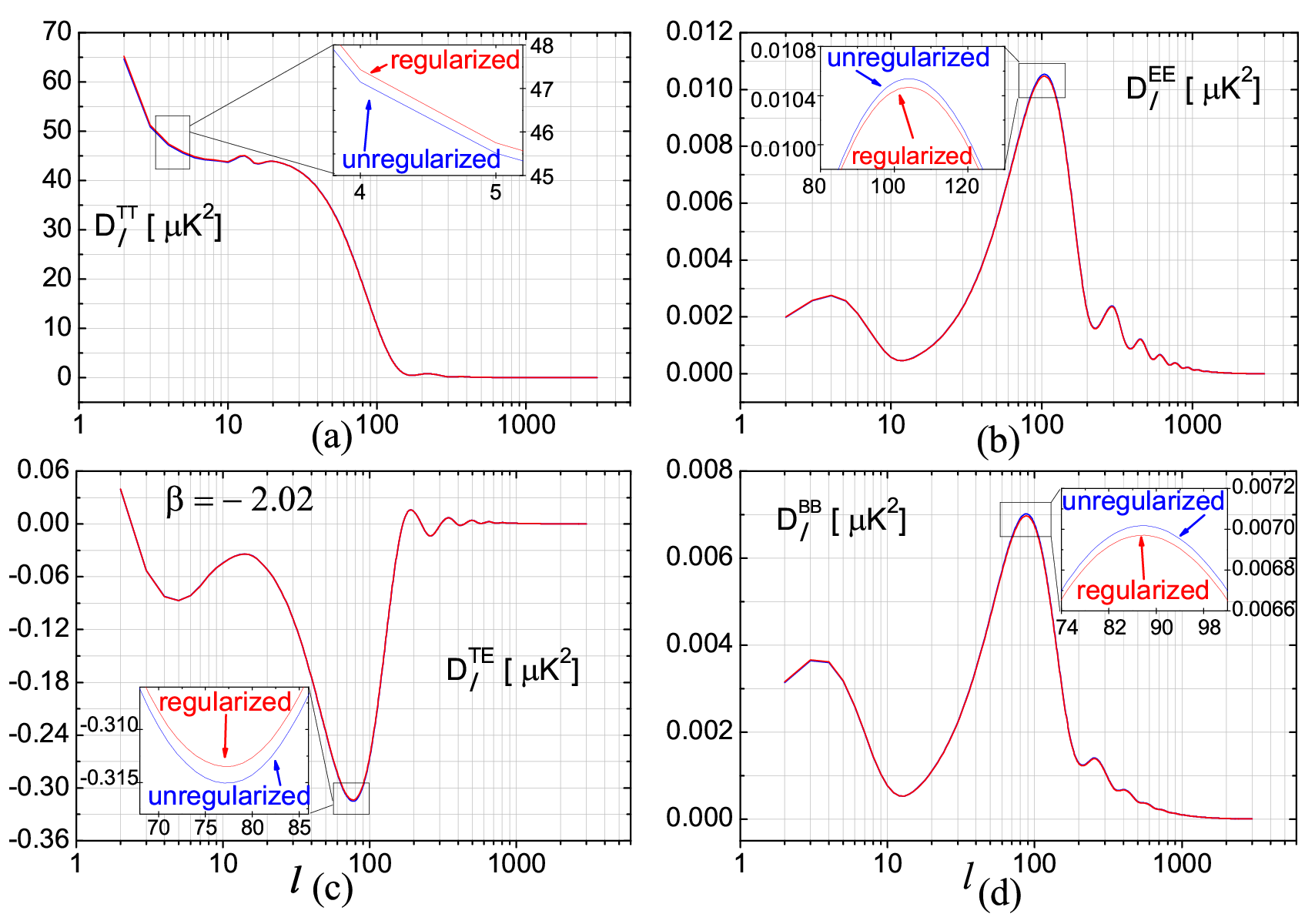}
\caption{ $l(l+1)C^{XX}_l/2\pi$ $[\mu K^2]$ generated by
    the regularized and
    unregularized power spectra of RGW,
    using CAMB code \cite{CAMBLewisChallinor2000A,CAMBLewisChallinor2000B}.
    The difference is tiny.
      }
     \label{CXXtensor}
\end{figure}

Given the  UV divergences of $\Delta_t^2 $,
we  subtract  them,
according to the minimal subtraction rule \cite{ParkerFulling1974,ParkerToms},
\be \label{regvac2count}
\Delta_t^2(k,\tau)_{ reg} =\frac{ A ^2k^3}{\pi^2 a^2(\tau)}
   \l( |v_{k}(\tau)|^2 - |v_k^{(2)}(\tau)|^2 \r)
~~~    \text{for all $k$},
\ee
which can apply at a time during inflation instantaneously.
We refer to this scheme as the all-$k$ regularization
since the subtraction  applies for all $k$-modes simultaneously.
$|v_k^{(2)}|^2$ is the counter part of 2nd adiabatic order
given by  (see (\ref{v2ex}) in Appendix A),
\be \label{counter2term}
|v_k^{(2)}|^2 = \frac{1}{2k} + \frac{ (\beta+1)\beta}{4k^3\tau^2},
\ee
where the first term   is to cancel
the quadratic divergence from the vacuum modes,
and the second term is to cancel the logarithmic divergence.
An important property  is that during inflation
the   spectrum  regularized at an earlier time
will evolve into a spectrum which is regularized at a later time,
as illustrated  in  Fig.\ref{4Rise} (b).
The  regularization  (\ref{regvac2count}) performed
at any time $\tau$ during inflation
always rightly removes the divergences which  rises up at $f_\tau$.
(This property is also  true
in the scheme of inside-horizon regularization in Section 4.)
The resulting spectrum  after regularization
is  $\propto k^{-2}$ at high frequency and becomes  UV convergent.
We show the  regularized and unregularized spectrum
at the end  of inflation $\tau_1$
in Fig.\ref{4Rise} (c) (d) for the models $\beta=-2.02$ and $\beta=-1.98$,
respectively.
In the range $(10^{-20} \sim 10^{-15})$Hz
the difference between the regularized and unregularized
is rather small ($\sim 0.2 \%$).
When they are substituted into the integration
$C^{XX}_l= \int_{k_{min}}^{k_{max}} \frac{dk}{k}\,  \Delta^2_t (k) P^2_{X\, l}(k)$
with $P^2_{X\, l}$ being the projection factors \cite{XiaZhang20089A,XiaZhang20089B},
they give their respective spectra of CMB  anisotropies and polarization.
We use CAMB code \cite{CAMBLewisChallinor2000A,CAMBLewisChallinor2000B,SeljakZaldarriaga1996}
and plot the resulting $C^{XX}_l$ in Fig.\ref{CXXtensor},
which show tiny differences.
(In computing  we have used
$k_{min}=7\times 10^{-8}$Mpc$^{-1}$ and $k_{max}=0.4$Mpc$^{-1}$,
which correspond to the present frequency
$f_{min} = 7\times 10^{-22}$Hz and $f_{max} =0.4\times 10^{-14}$Hz.
The lower and upper limits
of the numerical integration can only take finite values,
instead of $0$ and $\infty$,
so that the IR and UV divergences existing in $\Delta_t^2(k)$
are not reflected  in the numerical $C^{XX}_l$.)
Nevertheless,  in the range $(10^{-9} \sim 10^9)$Hz
the regularized power spectrum of RGW has been suppressed considerably
by a  factor as much as  $(1 \sim 10^3)$,
as seen in Fig.\ref{4Rise} (c) and (d).
This range covers the working bands of PTA \cite{SKA-Smits2009A,SKA-Smits2009B,
EPTA-Haasteren2011,Nan2011,NANOGrav-Demorest2013,Tong2014},
LISA  \cite{LISAwebA,LISAwebB} and LIGO \cite{LIGOweb}.
The suppression is an unwanted outcome of the  all-$k$  regularization.

\begin{figure}
\centering
\includegraphics[width=0.6\linewidth]{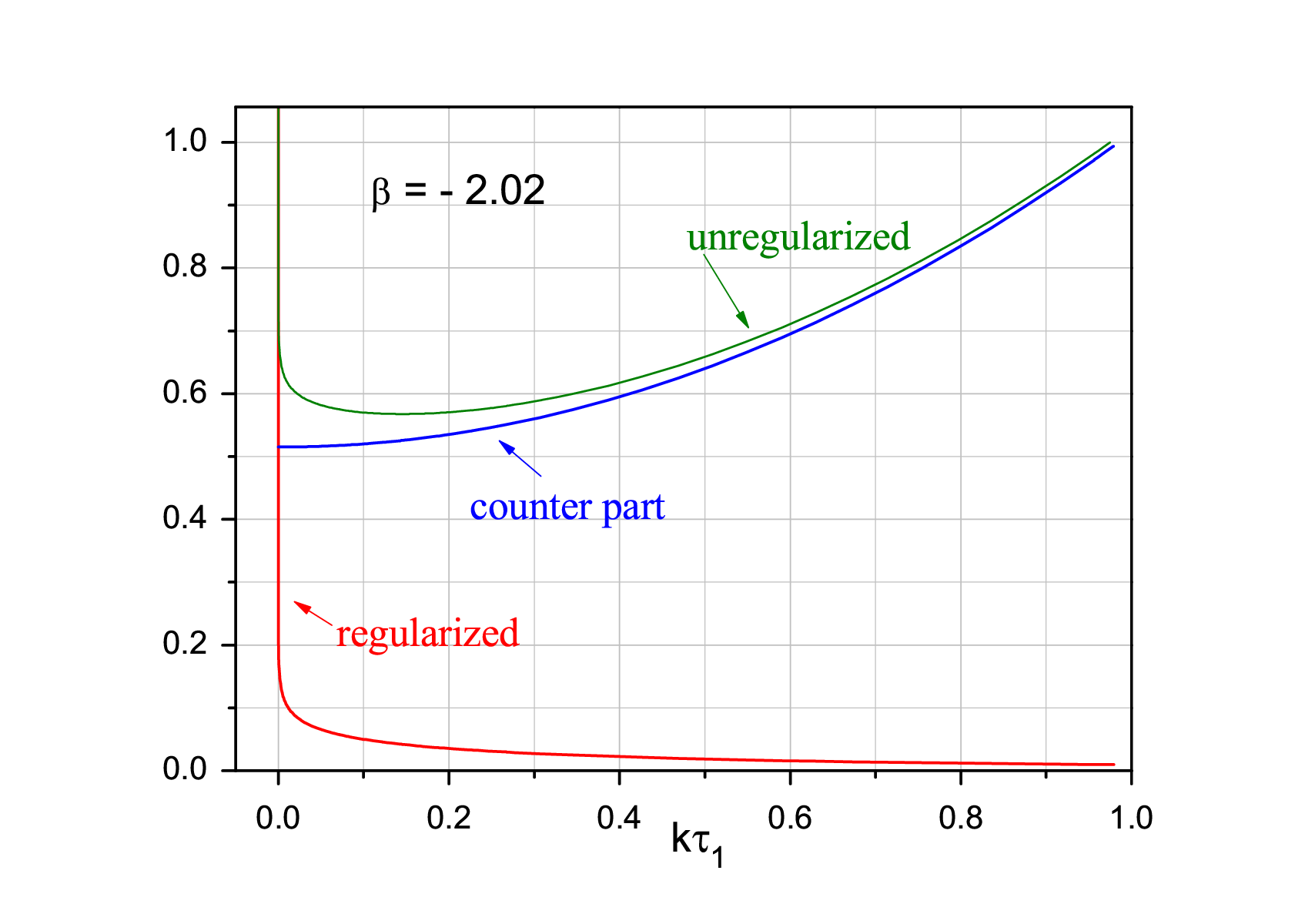}
\caption{ Schematically,
    Green: the unregularized spectrum  is IR divergent and dominant.
    Blue: the counter part is  IR log divergent and subdominant.
    Red:  the regularized   spectrum is IR divergent.
      }
     \label{unregulAndregul}
\end{figure}

\begin{figure}
\centering
\includegraphics[width=0.6\linewidth]{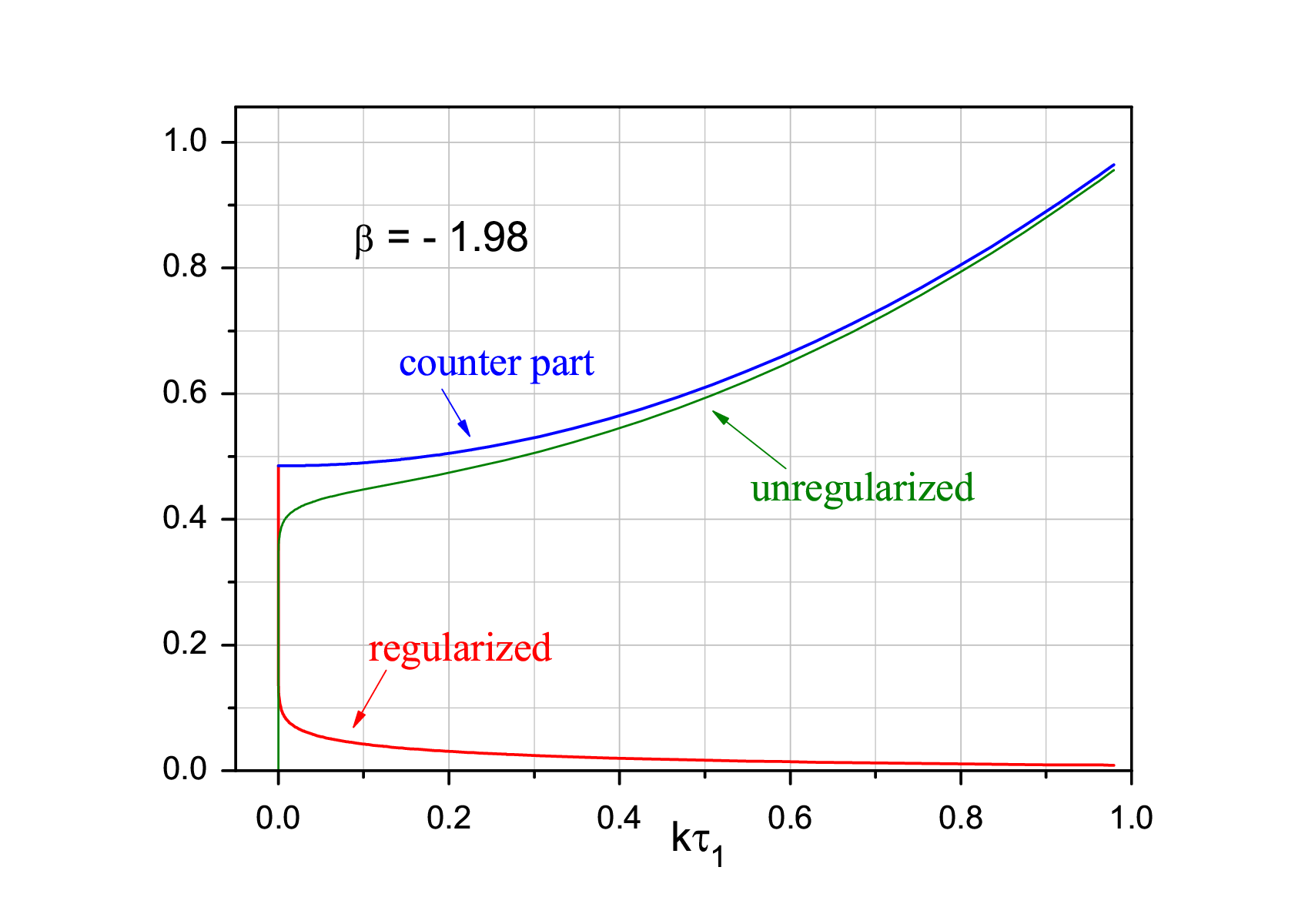}
\caption{ Schematically,
     Green: the unregularized   spectrum is IR convergent and subdominant.
    Blue: the counter part is  IR log divergent and dominant.
    Red:  the regularized  spectrum is  IR log divergent.
      }
     \label{unregulAndregul198}
\end{figure}

Let us examine the IR  behavior of $\Delta_t^2$ at  $k \sim 0$
under the regularization.
The counter part  at small $k$ gives
\be \label{counterlim}
k^3 \l(\frac{1}{2k} +\frac{ (\beta+1)\beta}{4k^3\tau_1^2} \r)
 \simeq
 \frac{ (\beta+1)\beta}{4  \tau_1 ^2},
\ee
which is a finite,  $k$-independent constant.
So, under the adiabatic subtraction,
the all-$k$  regularization  will downshift
$\Delta_t^2$  at $k \sim 0$  by this constant,
and consequently  the auto-correlation
will acquire an IR logarithmic divergent term as the following
\be\label{logdiv}
\frac{ (\beta+1)\beta}{4  \tau_1 ^2}
  \int_0 \frac{dk}{k}
=  \infty .
\ee
As said early,
for the models $\beta<-2$, the unregularized spectrum is already IR power divergent,
and the counter term is negligible in the limit $k|\tau|\ll 1$.
This case is schematically illustrated in  Fig.\ref{unregulAndregul},
and corresponds to
the situation discussed in Refs.\cite{UrakawaaStarobinsky2009,Alinea2018}
when the Hubble parameter at a later time during inflation
is smaller than the one at horizon-crossing, $H(\tau)/H(\tau_*)\ll 1$.
However, for the models  $\beta>-2$, i.e.,  $H(\tau)/H(\tau_*) > 1$,
which Refs.\cite{UrakawaaStarobinsky2009,Alinea2018} did not consider,
the unregularized spectrum is IR  convergent,
the counter term is IR logarithmic divergent,
so that the regularized spectrum is dominated
by the counter term and becomes IR logarithmic divergent,
as schematically illustrated in  Fig.\ref{unregulAndregul198}.
Therefore,
the all-$k$ regularization has a difficulty  for the $\beta>-2$ inflation models,
since it makes an IR convergent spectrum into an IR divergent one.
It also has a difficulty
with the   de Sitter inflation,
in which the unregularized spectrum (\ref{psbet2})
and the adiabatic counter part   (\ref{counter2term})
 cancel exactly,  yielding a vanishing power spectrum.

\begin{figure}
\centering
\includegraphics[width=0.6\linewidth]{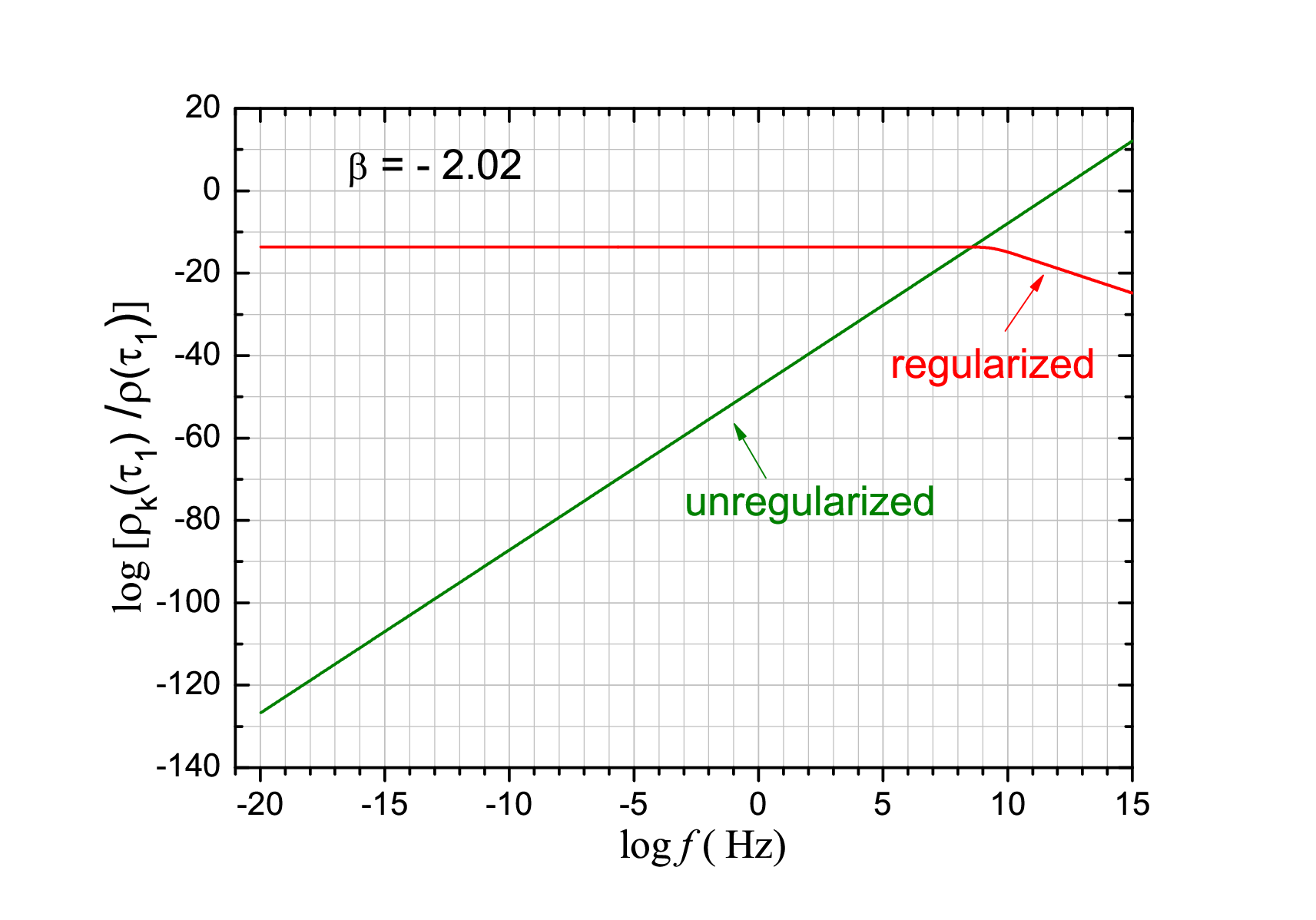}
\caption{  Red:  the spectral energy density by the all-$k$ regularization
            at   $\tau_1$.
         Green: the  unregularized.
     Rescaled by the background energy density $\rho(\tau_1)$ of inflation.
     }
     \label{energyCutInf2}
\end{figure}

\begin{figure}
\centering
\includegraphics[width=0.6\linewidth]{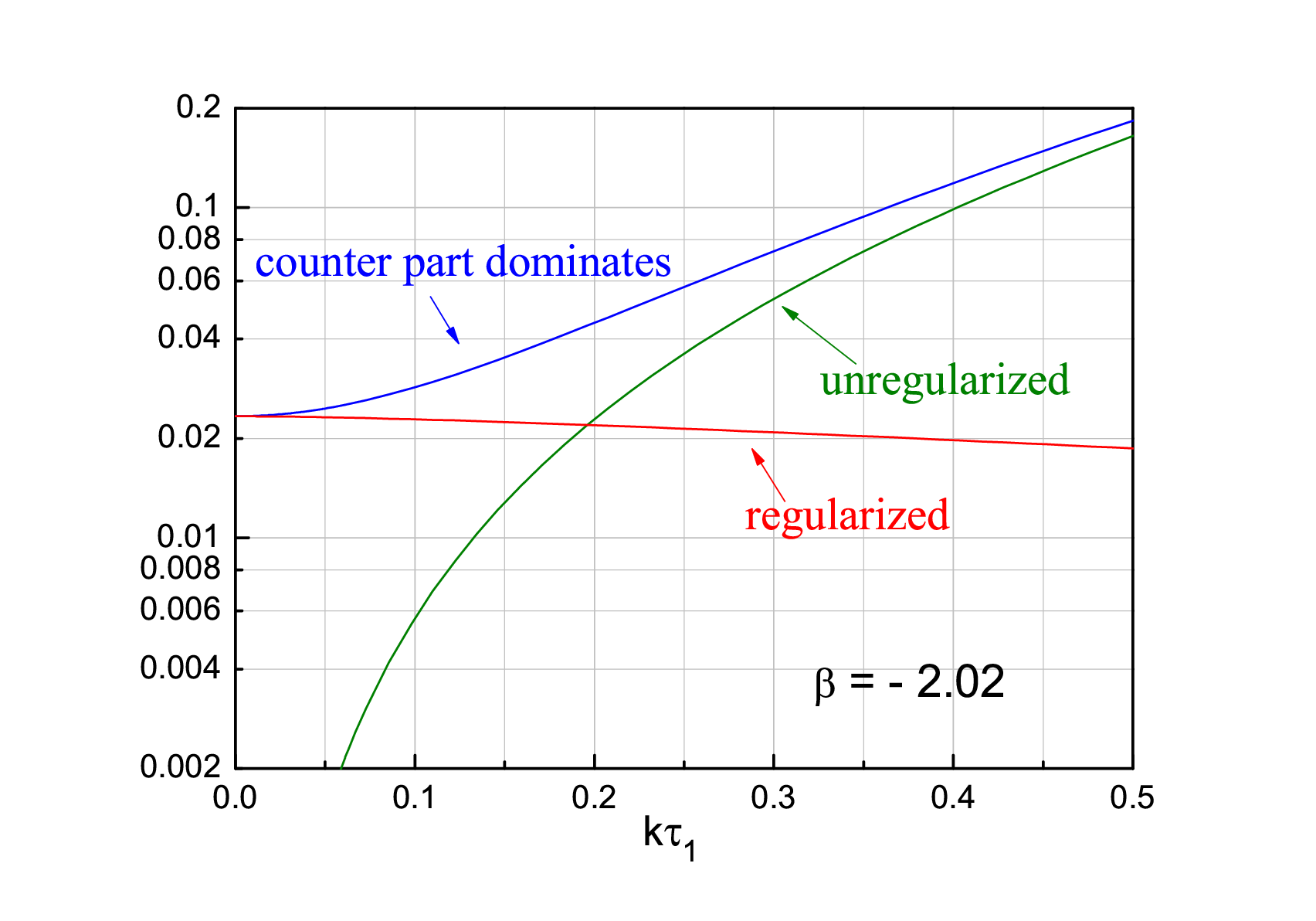}
\caption{ Schematically,
    Green: the unregularized pressure.
    Blue: the counter part dominates.
    Red: the regularized pressure.
        }
     \label{PunregulAndregul}
\end{figure}

More drastic low-frequency distortions
are brought about upon the energy density and pressure  by
the all-$k$ regularization.
The  energy density and pressure  in vacuum
contain UV quartic divergence of the 0th order ($\propto k^4$),
besides the quadratic   divergence of 2nd order ($\propto k^2$)
and logarithmic   divergence of 4th order ($\propto k^{0}$).
By the minimal subtraction rule,
the adiabatic regularization will be applied
to the 4th order,
\be\label{rhoreg1}
 \rho_k(\tau)_{ reg} = \frac{k^3}{\pi^2a^2}
             \Big|   | ( \frac{v_k(\tau)}{a} ) '  |^2
             -  | ( \frac{v_k^{(4)}(\tau) }{a} ) ' |^2  \Big|,
 ~~~ {\text{for all $k$}},
\ee
where  the 4th order  counter part during inflation  given by  (\ref{ad4v})
has the three terms,
\be \label{vprimi4}
 | ( \frac{v_k^{(4)}(\tau) }{a} ) ' |^2
   =\frac{1}{a^2}\left[\frac{k}{2}+\frac{(\beta+1)(\beta+2)}{4k\tau^2}
 +\frac{3\beta(\beta+1)(\beta+2)(\beta+3)}{16k^3\tau^4}\right],
\ee
which just cancels all the UV divergences of the squared derivative
$\left|\left( v_k / a \right) ' \right|^2$  in (\ref{expdvprim2}).
Similarly,  the  pressure is regularized by
\be\label{pressurereg1}
p_k(\tau)_{ reg}= \frac{k^5}{3\pi^2a^4}
\left| |v_k(\tau)|^2-|v_k^{(4)}|^2\right|
   ~~~ {\text{for all $k$}},
\ee
where the adiabatic 4th order counter part in  (\ref{4thorderterm2}) is
\be\label{4thorderterm}
 \left| v_k^{(4)}(\tau)    \right|^2  = \frac{1}{2 k}
    + \frac{\beta(\beta+1) }{4 k^3 \tau^2}
    + \frac{3(\beta -1 ) \beta (\beta+1) (\beta+2)}{16 k^5 \tau^4} .
\ee
Fig.\ref{energyCutInf2}
shows the resulting $\rho_k(\tau_1)_{ reg}$,
which is so drastically distorted that
it becomes flat at lower frequencies $f < 10^{9}$Hz.
Moreover, at the IR end $k\sim 0$,
it becomes IR logarithmic divergent.
The situation with pressure is similar.
This happens for all inflation models of $\beta \simeq -2$ other than  $\beta =-2$.
Recall that the original unregularized $\rho_k$ and $p_k$
in (\ref{presssmallk}) (\ref{enersmallk})  are IR convergent.
Let us analyze  the occurrence of IR divergence in details,
with the pressure as example,
by plotting each individual term  schematically
in   Fig.\ref{PunregulAndregul}.
The counter part (\ref{4thorderterm})  at small $k$
gives
\be \label{pconk}
   k^5 \Big( \frac{1}{2 k}
    + \frac{\beta(\beta+1) }{4 k^3 \tau_1^2}
    + \frac{3(\beta -1 ) \beta (\beta+1) (\beta+2)}{16 k^5 \tau_1^4} \Big)
       \simeq
      \frac{3(\beta -1 ) \beta (\beta+1) (\beta+2)}{16   \tau_1^4} ,
\ee
which is  a   $k$-independent constant.
Substituting  (\ref{pconk}) into the integration  (\ref{prefp})
will give a logarithmic divergent term
\[
\frac{3(\beta -1 ) \beta (\beta+1) (\beta+2)}{16   \tau_1^4}
   \int_0 \frac{dk}{k}
        =  \infty,
\]
so that $p_{gw}$ becomes IR logarithmic divergent.
Similar for the regularized energy density.
This  outcome at low $k$
is a difficulty for the scheme of all-$k$ regularization.


For de Sitter inflation,
the scheme has  another  difficulty.
The spectral energy density $\rho_k$ in (\ref{rhob2})
has only one quartic divergent term,
the counter terms
$|( \frac{v_k^{(4)} }{a} )'|^2$ in (\ref{vprimi4})
contain just one term $\frac{k}{2a^2}$,
so that they cancel,  yielding  a zero energy.
Similar for the spectral pressure $p_k$.

\section{Regularization inside the inflation  horizon}

We have seen in Fig.\ref{4Rise} (a)
that the unregularized power spectrum at a fixed time $\tau_1$
 rises up for $k \gtrsim 1/|\tau_1|$,
and the UV divergences come from this high $k$ range,
whereas the low  $k$ range is irrelevant to UV divergence.
This  observation   suggests that,
to remove UV divergences,
 we need only regularize the  $k$-modes
whose   wavelengths are  inside the inflation horizon ($k \gtrsim 1/|\tau_1|$),
and hold  intact the $k$-modes  outside the  horizon.

\begin{figure}
\centering
\includegraphics[width=0.6\linewidth]{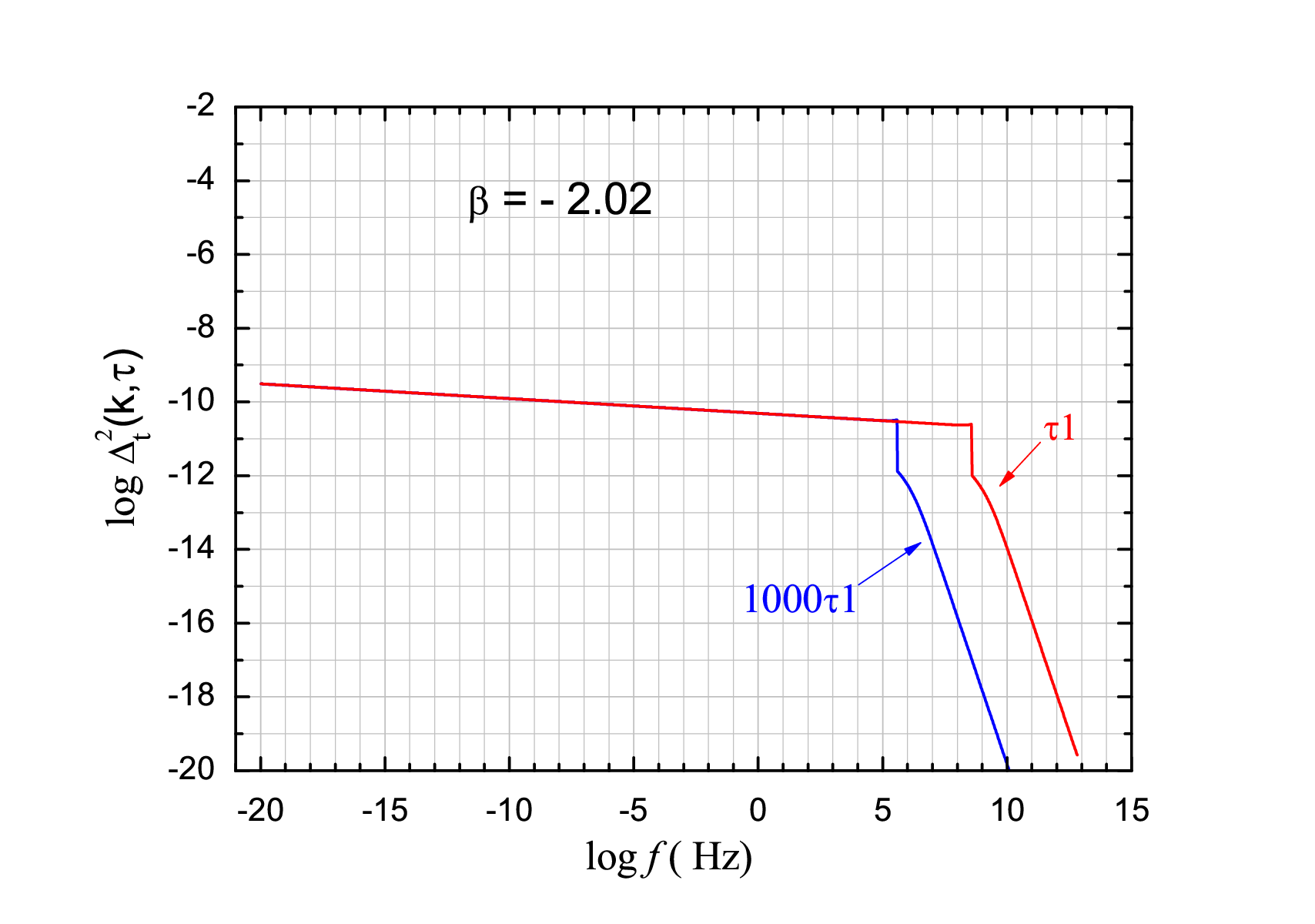}
\caption{
     The spectrum  regularized according to
     the inside-horizon regularization (\ref{regvac2counth2})
     at an earlier time $\tau=1000\tau_1$, and at  $\tau_1$,
     respectively.
     The former  will evolve into the latter  at $\tau_1$.
           }
     \label{RiseupRegInH1-1000}
\end{figure}

\begin{figure}
\centering
\includegraphics[width=0.6\linewidth]{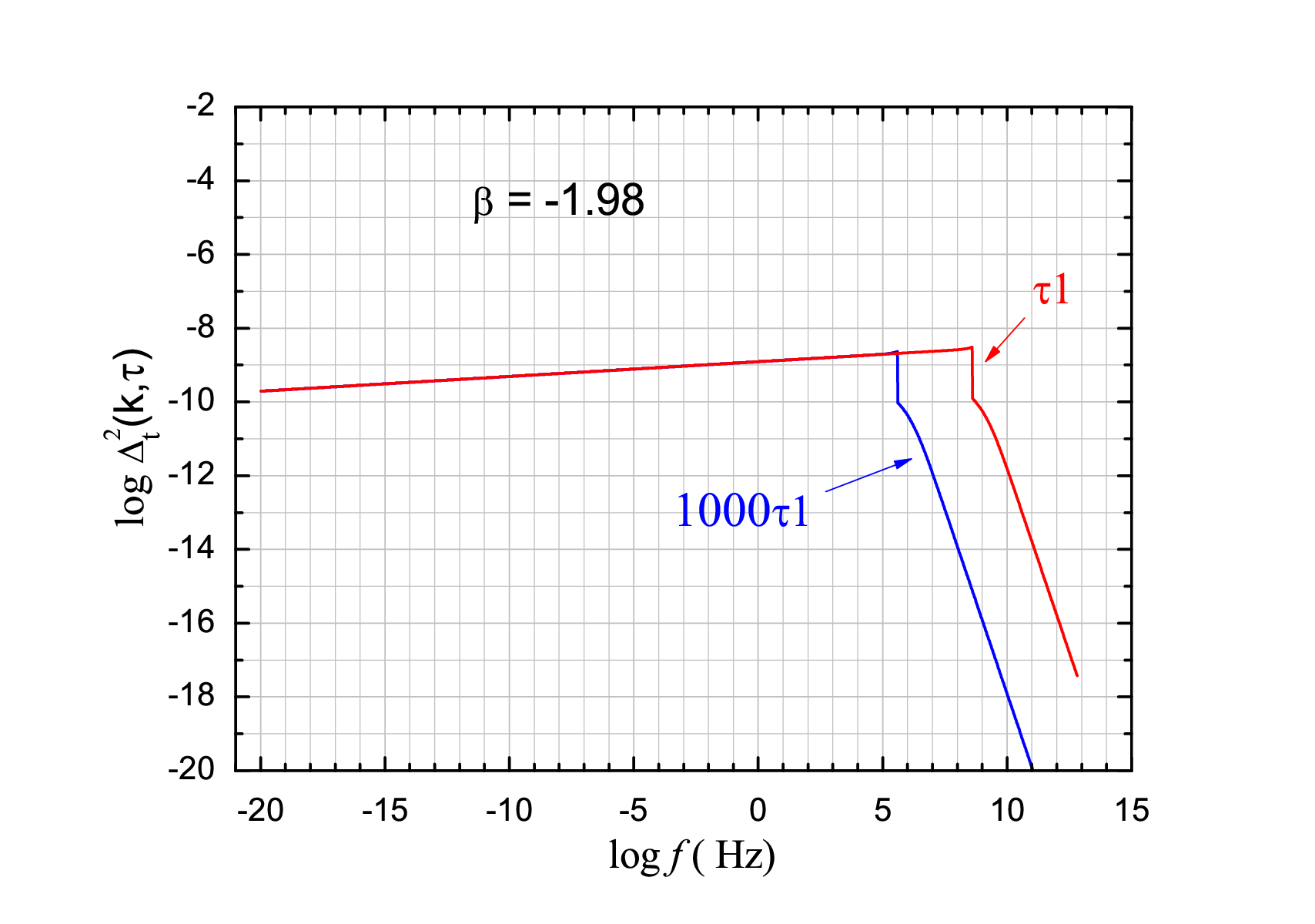}
\caption{ Similar to Fig.\ref{RiseupRegInH1-1000},
         for the model  $\beta = -1.98$.}
     \label{t1pwreg2198}
\end{figure}

\begin{figure}
\centering
\includegraphics[width=0.9\linewidth]{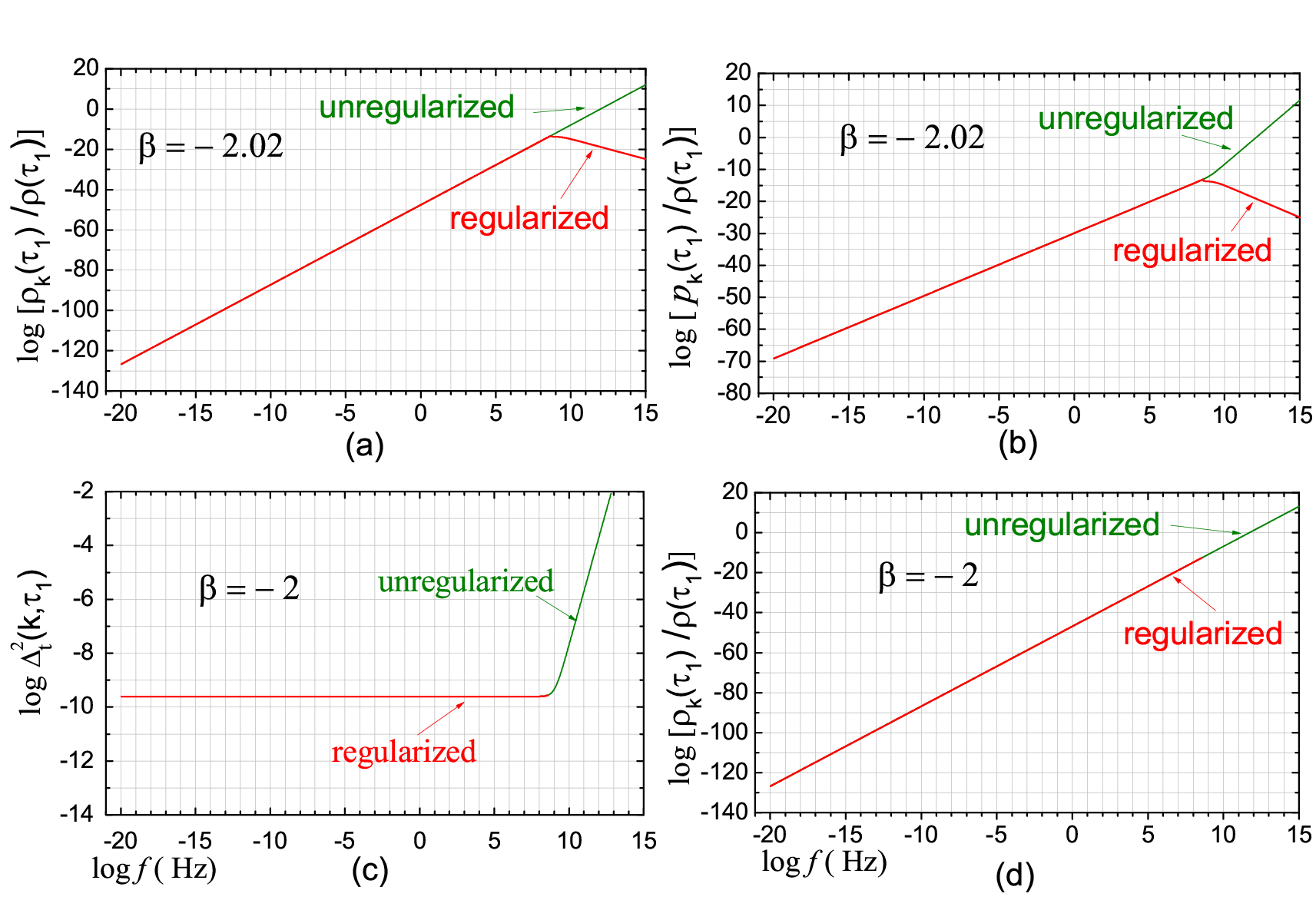}
\caption{
(a) $ \rho_k(\tau_1)_{reg}$
        by  Eq.(\ref{Hrhoreg}).
(b) $ p_k(\tau_1)_{ reg}$
        by Eq.(\ref{Hpressure}).
(c) $\Delta_t^2(k,\tau_1)_{reg} \ne 0$ for  $f<10^9$Hz
       by (\ref{regvac2counth2}) for de Sitter,
(d) $ \rho_k(\tau_1)_{reg} \ne 0$ for  $f<10^9$Hz
  by  Eq.(\ref{Hrhoreg})
     for de Sitter .
  }
     \label{infend4all}
\end{figure}

For the power spectrum
 we propose the following scheme of regularization
\ba \label{regvac2counth2}
  \Delta_t^2(k,\tau)_{reg} = \frac{A ^2  k^3}{\pi^2 a^2(\tau)}
\Bigg \{
 \begin{array}{ccc}
  &    ( |v_{k}|^2 - |v_k^{(2)}|^2 ) ,
  \,\,  & \text{ for $k \ge \frac{1}{|\tau_1|}$} ,
    \\
  &   |v_{k}|^2,
    \,\,\,\,\,  & \text{ for $k  <  \frac{1}{|\tau_1|}$} .
   \\
 \end{array}
\ea
The regularization is performed  instantaneously at  fixed $\tau_1$.
We refer to this scheme  also as the inflation-horizon regularization.
This can apply at any time $\tau_1$ during inflation,
the  spectrum  regularized at an earlier time
will evolve into the spectrum regularized at a later time.
All the resulting  spectra  regularized
at any instance   during inflation are equivalent,
in regard to post-inflation cosmology.
Once the initial spectrum is made  UV convergent,
it will continue to be so in  subsequent later stages.
The   power spectrum regularized
at $\tau=1000\tau_1$ and at $\tau_1$ is shown
in Fig.\ref{RiseupRegInH1-1000} for   $\beta = -2.02$,
and in Fig.\ref{t1pwreg2198} for  $\beta = -1.98$,  respectively.
As expected,
the low $k$ portion is intact,
whereas the UV divergences are gone,
yielding a spectrum  $ \propto k^{-2}$  at high frequency end.
For  de Sitter inflation,
the regularized power spectrum is nonvanishing  in the whole range $f< 10^{9}$Hz
and is zero for $f > 10^{9}$Hz,
as seen in Fig.\ref{infend4all} (c).
So  the difficulty of the all $k$ regularization
is overcome.

By the way,
for the scalar curvature perturbation
\cite{Sasaki1986,Hwang1994A,Hwang1994B,Gordon2000}
during inflation,
the power spectrum  has a similar structure to that of RGW \cite{WangZhangChen2016},
and can  be also regularized by the inside-horizon regularization
\be
\Delta^2_R (k,\tau)_{reg}
 =  \frac{\beta+1}{16(\beta+2)} \Delta^2_t (k,\tau)_{reg}
\ee
for  $\beta\ne -2$
in the exponential inflation model of  Eq.(\ref{inflation}).
We mention that
the CMB spectra $C^{XX}_l$
induced by $\Delta^2_R (k,\tau)_{reg}$
are the same as those by
the unregularized  spectrum $\Delta^2_R (k,\tau)$,
since its low $k$ portion is unchanged under
the inside-horizon regularization.


For $\rho$ and $p$, parallel to (\ref{regvac2counth2}),
we have the inside-horizon regularization
to  4th adiabatic order as the following
\ba \label{Hrhoreg}
 \rho_k(\tau)_{reg}  = \frac{  k^3}{\pi^2 a^2(\tau)}
\Bigg \{
 \begin{array}{ccc}
  &   \Big| | ( \frac{v_k }{a} ) '  |^2
             -  | ( \frac{v_k^{(4)}  }{a} ) ' |^2  \Big | ,
  \,\,  & \text{ for $k \ge \frac{1}{|\tau_1|}$} ,
    \\
  &   | ( \frac{v_k }{a} ) '  |^2,
    \,\,\,\,\,  & \text{ for $k < \frac{1}{|\tau_1|}$} ,
   \\
 \end{array}
\ea
\ba \label{Hpressure}
 p_k(\tau)_{reg}  = \frac{  k^5}{3\pi^2 a^4(\tau)}
\Bigg \{
 \begin{array}{ccc}
  &  \Big| |v_k |^2-|v_k^{(4)}|^2  \Big|  ,
  \,\,  & \text{ for $k \ge \frac{1}{|\tau_1|}$} ,
    \\
  &    |v_k |^2,
    \,\,\,\,\,  & \text{ for $k < \frac{1}{|\tau_1|}$} .
   \\
 \end{array}
\ea
The resulting  regularized  spectra are plotted
in Fig.\ref{infend4all} (a) and  (b),
respectively.
For  de Sitter inflation,
$ \rho_{k\, reg} $   is non-vanishing  for $f< 10^{9}$Hz
and is zero for  $f > 10^{9}$Hz,
as shown in  Fig.\ref{infend4all} (d).
And  $p_{k\, reg} $ is  similar.
Thus,  the inside-horizon scheme   overcomes
the difficulty of the all-$k$ scheme  for de Sitter inflation.

At the level of the linearized approximation of Einstein equation,
RGW is a linear field,  and its $k$-modes are independent each other,
there is no energy exchange between different  $k$-modes.
Thus,
the regularization inside the horizon  will not affect the those outside the horizon.
For the outside-horizon $k$-modes,
$t_{\mu\nu}$  combined with other matter components
is originally conserved
\cite{Isaacson1968A,Isaacson1968B,Abramo1997,Giovannini2006,SuZhang}.
The procedure of  (\ref{Hrhoreg}) (\ref{Hpressure})
is equivalent to inserting
a step function $\theta(k|\tau_1|-1)$ in front of the subtraction term,
where $\tau_1$ is fixed, not a time variable.
So time differentiation just passes over $\theta(k|\tau_1|-1)$
which will not cause violation of conservation.
  Under the inside-horizon scheme
the covariant conservation is respected
by RGW together with other matter components.
The scheme of inside-horizon regularization
can be applied to other linear fields during inflation,
such as  free scalar fields, either massless or massive,
as long as interaction between $k$-modes is vanishing,
 or negligibly small.
This is the case also for
the scalar curvature perturbation \cite{Sasaki1986,WangZhangChen2016}
and the gauge-invariant perturbed scalar field \cite{Hwang1994A,Hwang1994B,Gordon2000},
in the exponential inflation model of  Eq.(\ref{inflation}).
As for the second order perturbation,
there will be coupling between different $k$-modes of RGW,
and between RGW and scalar metric perturbation \cite{WangZhang2017A,WangZhang2017B},
this scheme may not apply directly.

The regularized energy density of RGW is finite,
and  satisfies the so-called back-reaction constraint.
To be specific,   at time $\tau_1$,
we  substitute   $\rho_{k}(\tau_1)_{reg}$ into
the integration  (\ref{energyspectr})
to get the regularized  energy density
\be \label{energafreg}
\rho_{gw\, reg}
= \int _0 ^{k_1}   \rho_k(\tau_1)_{reg} \frac{d k}{k}
+  \int _{k_1} ^{\infty}    \rho_k(\tau_1)_{reg} \frac{d k}{k} ,
\ee
where $k_1 \equiv  1/|\tau_1|$.
The first term of (\ref{energafreg}) is an integration over modes outside the horizon,
 $\rho_k(\tau_1)$ can be  approximated by $\frac{k^4}{2 \pi^2a^4(\tau_1)}$ of (\ref{rhob2}),
so that
\be
\int _0 ^{k_1}  \rho_k(\tau_1)_{reg} \frac{d k}{k}
 \simeq  \frac{H^4  }{8 \pi^2 }.
\ee
In the second term of (\ref{energafreg}),
 $\rho_k(\tau_1)_{reg}$
has been given by (93) in Ref.\cite{WangZhangChen2016},
so that
$\int _{k_1} ^{\infty}   \, \rho_k(\tau_1)_{reg} \frac{d k}{k}
 \sim    \frac{(\beta+2)}{\pi^2} H^4 \ll H^4 $.
Hence,
the sum is $\rho_{gw\, reg} \simeq \frac{H^4  }{8 \pi^2 } $,
which is much smaller than the background energy density
$\frac{3}{8\pi G} H^2 $.

\section{Evolution of regularized spectra into the present}

\begin{figure}
\centering
\includegraphics[width=0.7\linewidth]{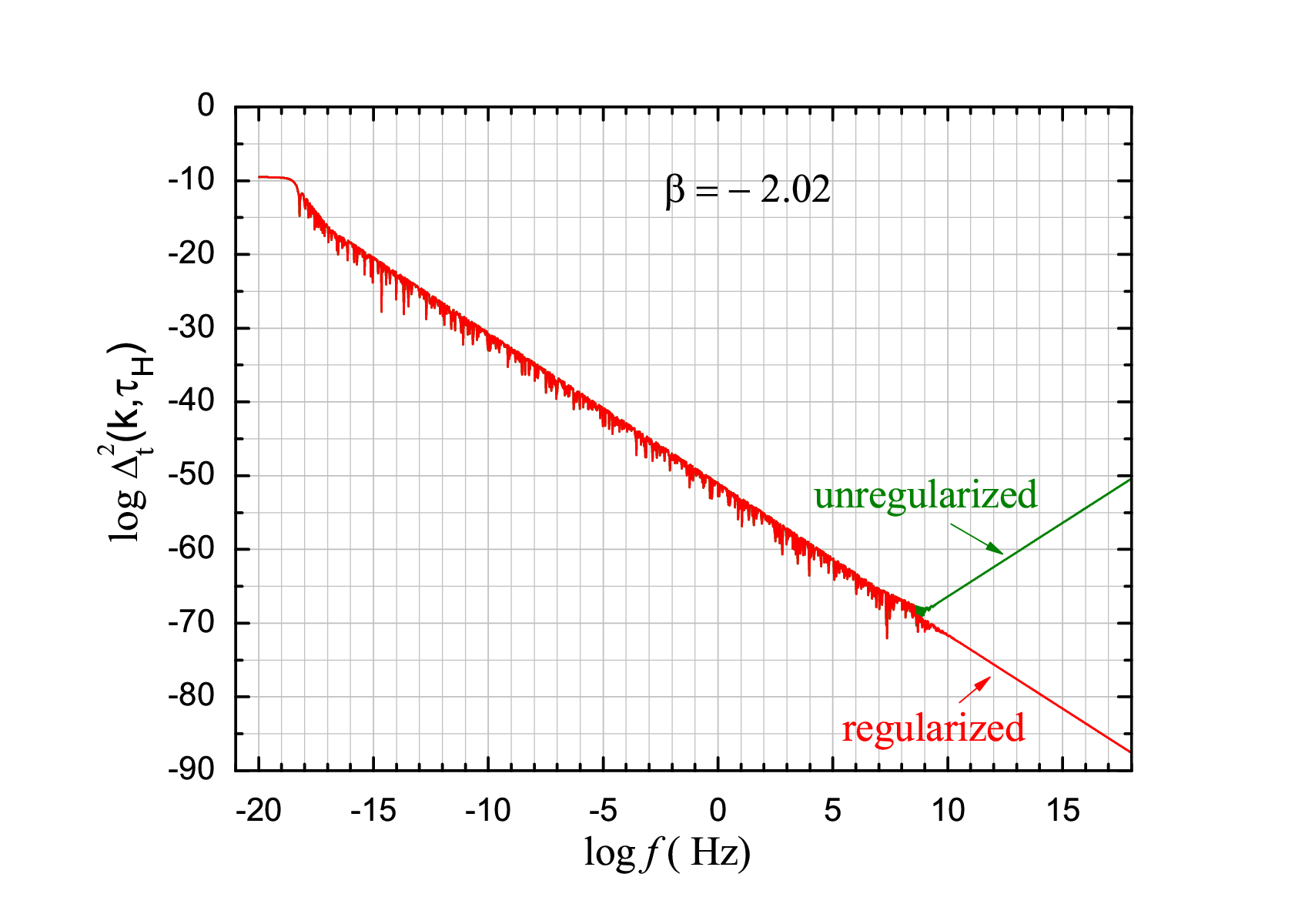}
\caption{  The present $\Delta_t^2(k,\tau_H)_{reg}$
which has evolved from the initial $\Delta_t^2(k,\tau_1)_{reg}$
     in Fig.\ref{RiseupRegInH1-1000}.
Its  low frequency portion  $f< 10^9$Hz
 is  overlapping with the unregularized.
   }
     \label{specInfToAcc2}
\end{figure}

\begin{figure}
\centering
\includegraphics[width=0.7\linewidth]{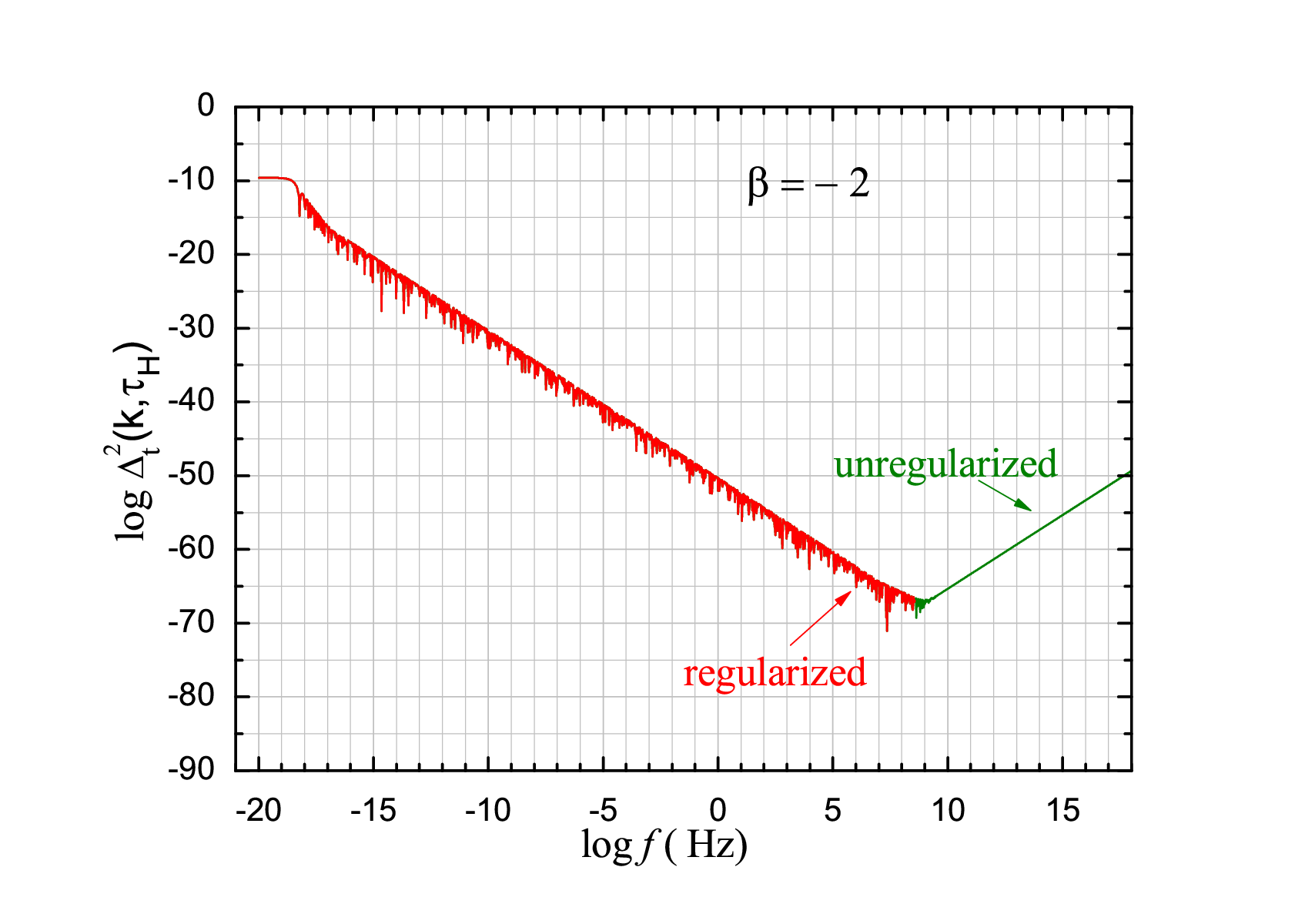}
\caption{  The present $\Delta_t^2(k,\tau_H)_{reg}$
   is  nonvanishing in the  range  $f<10^{9}$Hz,
   which has evolved from the initial $\Delta_t^2(k,\tau_1)_{reg}$
   in Fig.\ref{infend4all} (c).
    }
     \label{specInfToAccbeta2}
\end{figure}

\begin{figure}
\centering
\includegraphics[width=0.7\linewidth]{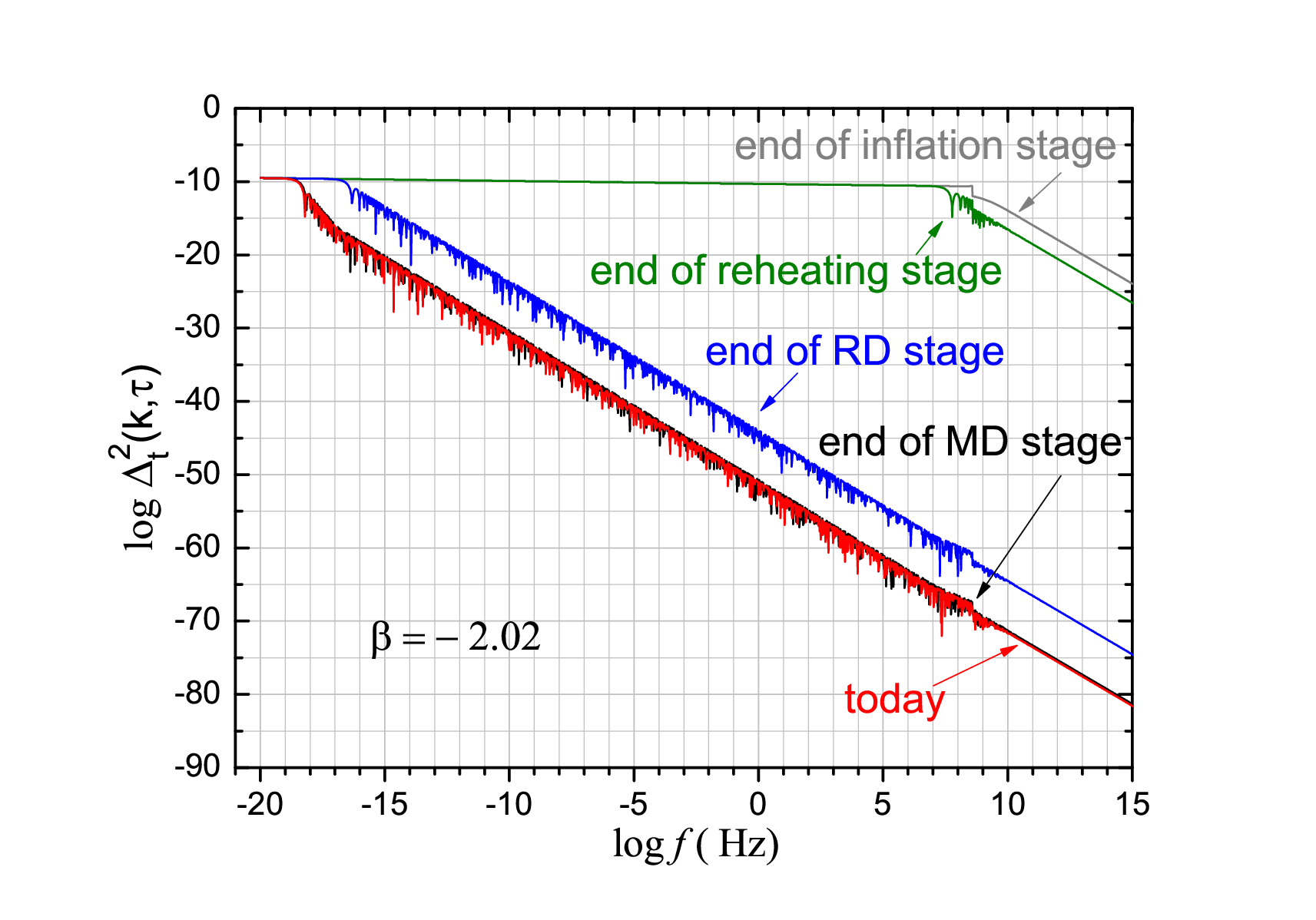}
\caption{ The evolution history of $\Delta^2_t(k,\tau)_{reg}$,
from inflation, reheating, radiation,  matter,
up to the present accelerating stage.
  }
     \label{evolpower5}
\end{figure}

Taking  the power spectrum by  (\ref{regvac2counth2})
as the initial condition,
we let it evolve to  the present stage.
Inside the horizon $k|\tau_1|>1$ the initial modes are taken to be
\ba\label{ukReInftoAcc2}
u_k^{reg}(\tau_1) = e^{i\theta(\tau_1)}
\sqrt{  |v_{k}(\tau_1)|^2 - |v_k^{(2)}(\tau_1)|^2  } \ ,
       \,\,\,   \text{for $k > \frac{1}{|\tau_1|}$}
\ea
with  the   phase  $e^{i\theta(\tau_1)}=\frac{v_k(\tau_1)}{|v_k(\tau_1)|}$.
Outside  the horizon,
the initial modes are the original $v_k(\tau_1)$.
The subsequent evolution of each regularized mode is independent,
and goes
through reheating, radiation, matter, up to the present accelerating stage.
Specifically,
using the mode (\ref{ukReInftoAcc2}) and its time derivative
of the inflation stage,
we calculate the coefficients $b_1, b_2$ for the reheating stage,
then for the subsequent stages  (see Appendix C).
This procedure  results in a regularized mode $u_k(\tau_H)_{reg}$
at the present time $\tau_H$,
and the associated regularized power spectrum
\be\label{psevlH}
\Delta^2_t(k,\tau_H)_{reg}
  = \frac{ k^{3}}{\pi^2a^2(\tau_H)}\frac{4}{M_{Pl}^2}|u_k(\tau_H)_{reg}|^2 .
\ee
Here the mode $u_k(\tau_H)_{reg}$  of each $k$  is actually
a combination of positive and negative frequency $k$-modes.
The  spectrum is shown in Fig. \ref{specInfToAcc2} for $\beta = - 2.02$,
and in  Fig.\ref{specInfToAccbeta2} for  $\beta = - 2$
which is nonvanishing for $f<10^{9}$Hz.
The detailed evolution history of regularized power spectrum
in the course of cosmic expansion
is demonstrated in Fig.\ref{evolpower5}.
The inside-horizon modes are decreasing as $h_k(\tau) \propto 1/a(\tau)$,
and the outside-horizon modes  keep constant $h_k(\tau) \propto$ const,
so that  higher-$k$ modes started decreasing earlier and has dropped more.

\begin{figure}
\centering
\includegraphics[width=0.9\linewidth]{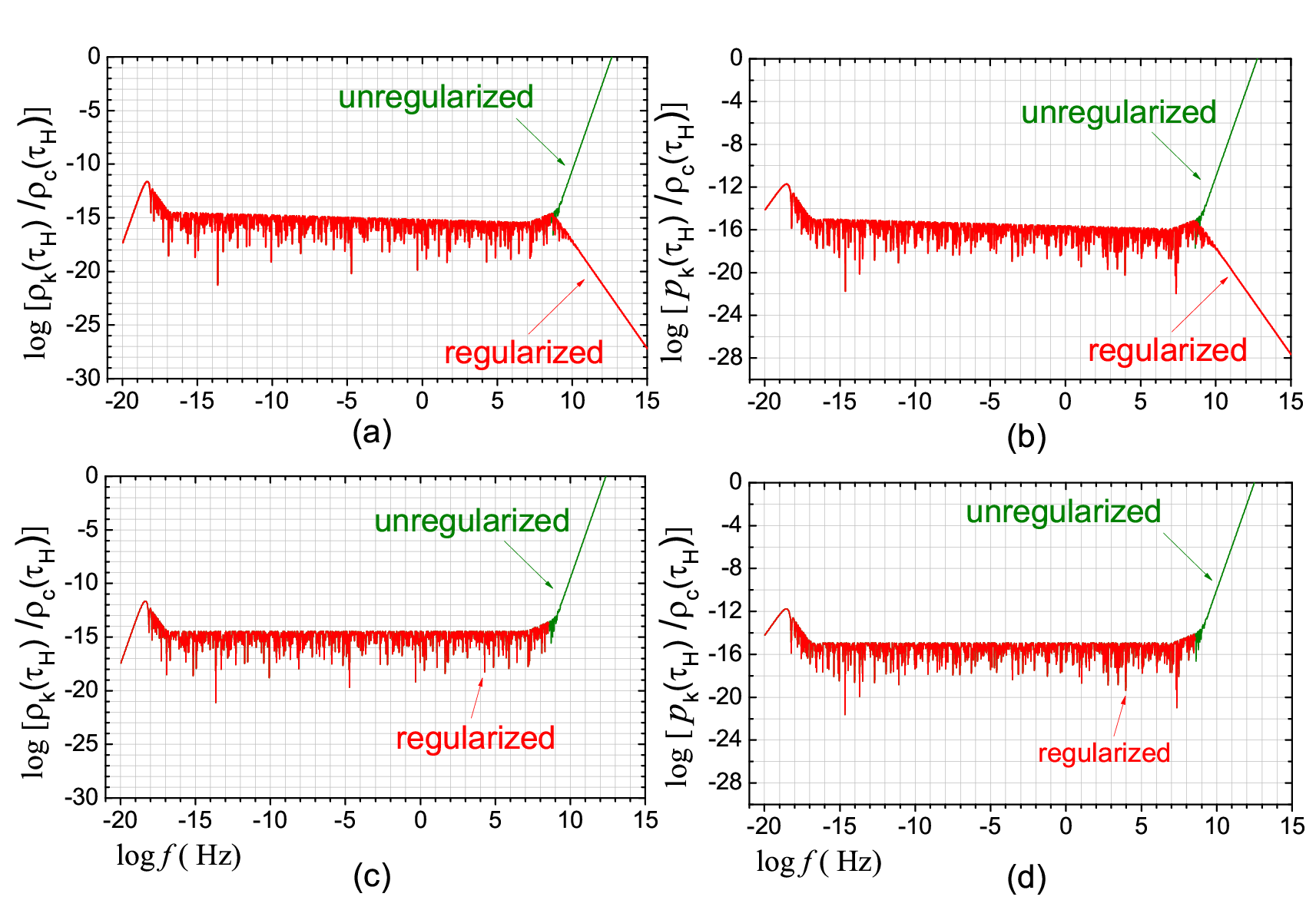}
\caption{
(a)   $\Omega_g(\tau_H)=\rho_k(\tau_H)/\rho_c$
  evolved from the initial
   $\rho_k(\tau_1)_{reg}$ in Fig.\ref{infend4all} (a).
(b)  $p_k(\tau_H)/\rho_c$
     evolved from  $p_k(\tau_1)_{reg}$
       in Fig.\ref{infend4all} (b).
(c) (d) For de Sitter inflation, $\Omega_g(\tau_H) \ne 0$,
   $p_k(\tau_H)/\rho_c \ne 0$
   for $f< 10^9$Hz.
         }
     \label{rhopressure4}
\end{figure}

\begin{figure}
\centering
\includegraphics[width=0.7\linewidth]{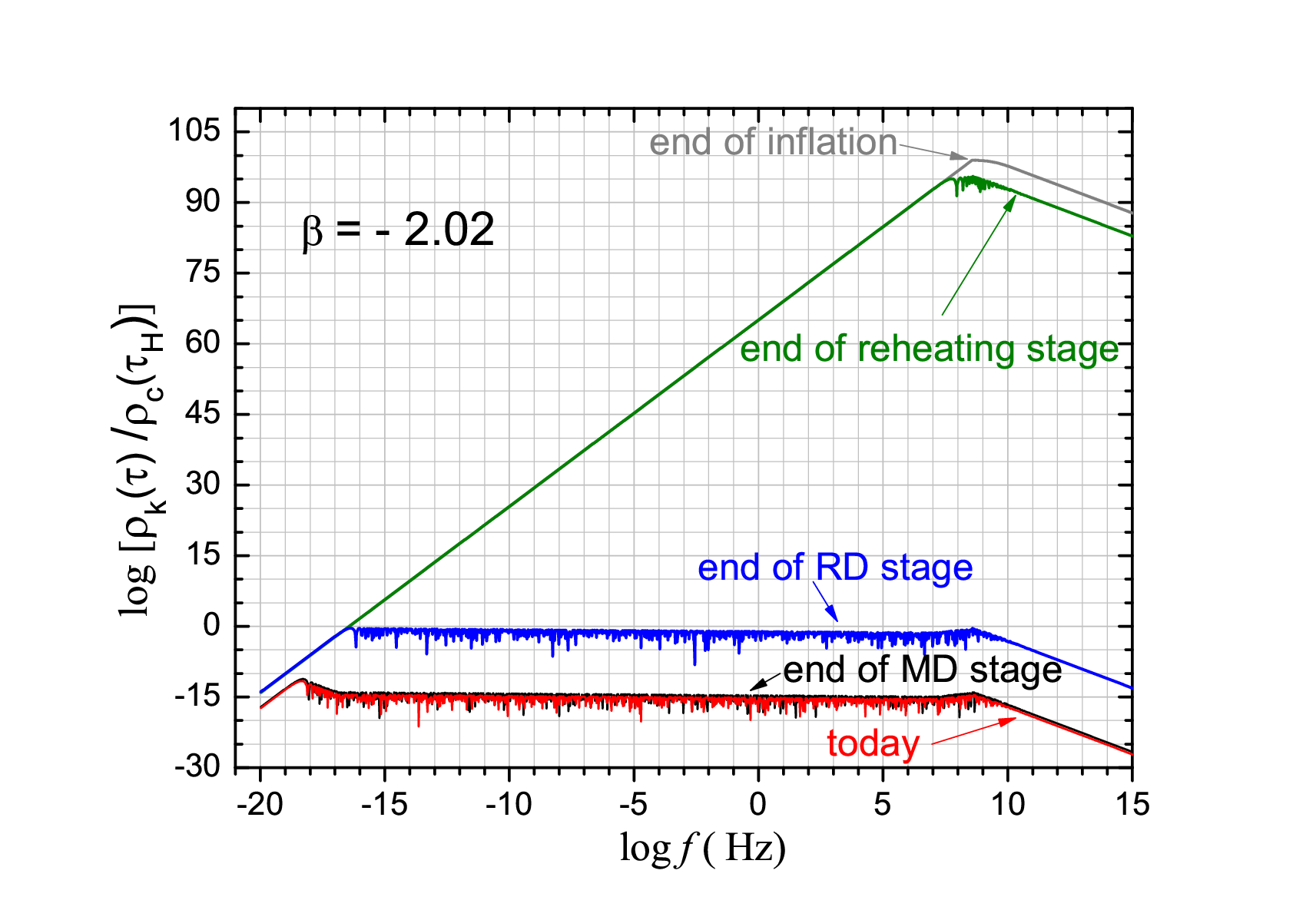}
\caption{  The evolution history of $ \rho_k(\tau)_{reg}$
   from the end of inflation up to the present.
         }
     \label{evolregenergy5}
\end{figure}

Now the evolution of $\rho$ and $p$.
The initial modes inside the   horizon are taken to be
\ba \label{initialvcut}
v_k^{reg}(\tau_1)
&=&
e^{i\theta(\tau_1)}\sqrt{
   \left| |v_{k}(\tau_1)|^2 - |v_k^{(4)}(\tau_1)|^2 \right| } ,
 \,\,\,  \ \text{ for $k|\tau_1|>1$}   \ ,
\ea
where  $e^{i\theta(\tau_1)}=\frac{v_k(\tau_1)}{|v_k(\tau_1)|}$,
and the  outside-horizon initial modes are the original $v_k(\tau_1)$.
The evolution
results in the regularized  modes   $u_k(\tau_H)_{reg}$  at the present time,
and the associated regularized spectral  energy density and pressure
\be
\rho_k(\tau_H)_{reg} = \frac{k^3}{\pi^2a^2(\tau_H)}
 \left| ( \frac{u_k(\tau_H)_{reg}}{a} ) '  \right|^2 ,
\ee
\be
p_k(\tau_H)_{ reg}= \frac{k^5}{3\pi^2a^4(\tau_H)}
   \left |u_k(\tau_H)_{reg} \right|^2 ,
\ee
$\rho_k(\tau_H)_{reg}$ is shown in Fig.\ref{rhopressure4} (a) for $\beta=-2.02$
and  Fig.\ref{rhopressure4} (c) for $\beta=-2$,
$p_k(\tau_H)_{reg} $ is shown in Fig.\ref{rhopressure4} (b) for $\beta=-2.02$
and Fig.\ref{rhopressure4} (d) for $\beta=-2$.
They are rescaled by the   critical density $\rho_c$ in plotting.
The detailed evolution history of
the regularized spectral energy density
is demonstrated in Fig.\ref{evolregenergy5}.

We notice that the spectra after inflation all exhibit quick oscillations
 in frequency domain,
whereas the initial spectra defined in BD vacuum during inflation have no oscillations.
The oscillatory pattern is  produced
in the consecutive expansion stages,
and is not changed by adiabatic regularization.

\section{ Structure of RGW¡¡and Interference }

\begin{figure}
\centering
\includegraphics[width=0.7\linewidth]{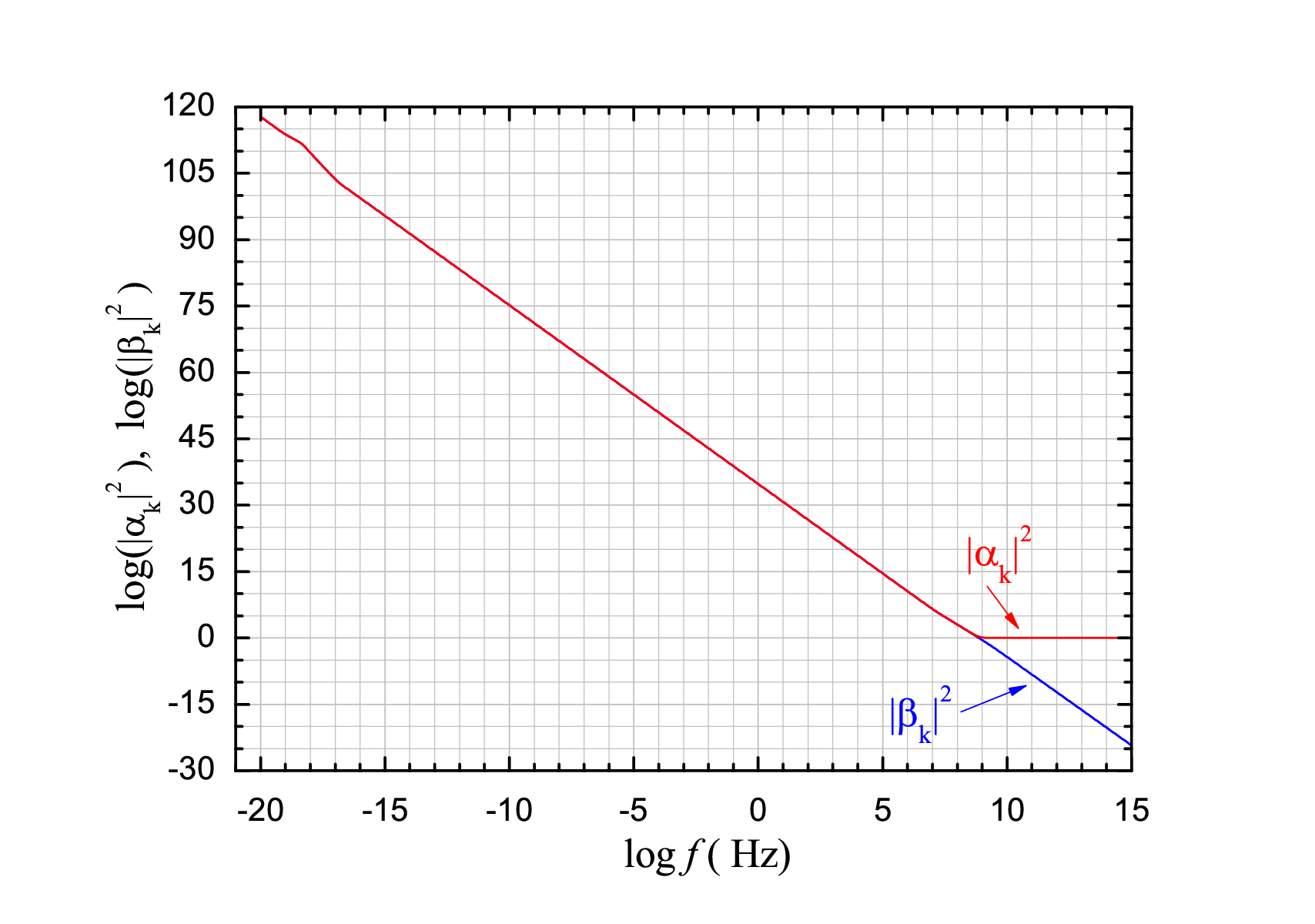}
\caption{  Blue:  $|\alpha_k|^2$.
  Red:  $|\beta_k|^2$.
The relation  $|\alpha_k|^2-|\beta_k|^2 =1 $ holds.}
     \label{alphakbetak}
\end{figure}

We  analyze the structure of RGW as a quantum field
in  the present accelerating  stage
with the scale factor  $a(\tau)=l_H|\tau-\tau_a|^{-\gamma}$
and   $\gamma \simeq 2.1$.
The solution   (\ref{solgen}) in the present stage is
\be \label{upresent}
u_k(\tau ) = \sqrt{\frac{\pi}{2}}\sqrt{\frac{s}{2k}}
     \big[e^{-i\pi\gamma/2}\beta_k  H^{(1)}_{-\gamma-\frac{1}{2} } (s)
          +e^{i\pi\gamma/2}\alpha_k  H^{(2)}_{-\gamma-\frac{1}{2}} (s) \big],
         \,\,\,\,\, \, \tau_E <\tau\leq \tau_H   ,
\ee
consisting of both positive and negative  modes,
where $s=k|\tau-\tau_a|$,
the Bogolyubov  coefficients $\beta_k, \alpha_k$ are
  determined by the cosmic evolution from initial condition.
Their analytical expressions are obtained
by connecting the modes $u_k$ and  $u_k'$ for each $k$
consecutively for the five stages.
See Appendix C and Ref.\cite{Zhang06A,Zhang06B}.
At high frequencies,
they  have the following asymptotic expressions:
{\allowdisplaybreaks
\ba\label{e1}
\beta_k&=&
-\frac{i}{2 z_2^2}
 e^{i (-x_1-t_1+t_s-y_s+y_2+z_2-z_E-s_E)+{\frac12  i \pi  \beta } }
\nn\\
&&
+i \left(\frac{1}{2 z_E^2}
-\frac{\gamma  (\gamma +1)}{4 s_E^2}\right)
e^{i (-x_1-t_1+t_s-y_s+y_2-z_2+z_E-s_E)+{\frac12  i \pi  \beta } }
\nn\\
&&
+i \left(\frac{\beta  (\beta +1)}{4 x_1^2}
-\frac{\beta_s (\beta_s+1)}{4 t_1^2}\right)
e^{i (-x_1+t_1-t_s+y_s-y_2+z_2-z_E-s_E)+{\frac12  i \pi  \beta } }
\nn\\
&&
+i\frac{ \beta_s (\beta_s+1) }{4 t_s^2}
e^{i (-x_1-t_1+t_s+y_s-y_2+z_2-z_E-s_E)+{\frac12 i \pi  \beta } }
+\mathcal O\left(k^{-3}\right) ,
\ea
\ba\label{e2}
\alpha_k&=&
i\bigg(
1
-i\frac{\beta  (\beta +1)}{2 x_1}
-i\frac{\beta_s (\beta_s+1)}{2 t_1}
+i\frac{\beta_s (\beta_s+1)}{2 t_s}
-i\frac{1}{z_2}
+i\frac{1}{z_E}
+i\frac{\gamma  (\gamma +1)}{2 s_E}
\nn\\
&&
-\frac{ \beta ^2 (\beta +1)^2}{8 x_1^2}
-\frac{ \beta_s^2 (\beta_s+1)^2}{8 t_1^2}
-\frac{ \beta_s^2 (\beta_s+1)^2}{8 t_s^2}
-\frac{1}{2 z_2^2}
-\frac{1}{2 z_E^2}
-\frac{ \gamma ^2 (\gamma +1)^2}{8 s_E^2}
\nn\\
&&
-\frac{ \beta  (\beta +1) \beta_s (\beta_s+1)}{4 x_1 t_1}
+\frac{ \beta  (\beta +1) \beta_s (\beta_s+1)}{4 x_1 t_s}
-\frac{ \beta  (\beta +1)}{2 x_1 z_2}
+\frac{ \beta  (\beta +1)}{2 x_1 z_E}
\nn\\
&&
+\frac{ \beta  (\beta +1) \gamma  (\gamma +1)}{4 x_1 s_E}
+\frac{ \beta_s^2 (\beta_s+1)^2}{4 t_1 t_s}
-\frac{ \beta_s (\beta_s+1)}{2 t_1 z_2}
+\frac{ \beta_s (\beta_s+1)}{2 t_1 z_E}
\nn\\
&&
+\frac{ \beta_s (\beta_s+1) \gamma  (\gamma +1)}{4 t_1 s_E}
+\frac{ \beta_s (\beta_s+1)}{2 t_s z_2}
-\frac{ \beta_s (\beta_s+1)}{2 t_s z_E}
\nn\\
&&
-\frac{ \beta_s (\beta_s+1) \gamma  (\gamma +1)}{4 t_s s_E}
+\frac{1}{z_2 z_E}
+\frac{ \gamma  (\gamma +1)}{2 z_2 s_E} \nn \\
&& -\frac{ \gamma  (\gamma +1)}{2 z_E s_E}
\bigg)
e^{i (-x_1-t_1+t_s-y_s+y_2-z_2+z_E+s_E)+ \frac12 i \pi  \beta}
+\mathcal O\left(k^{-3}\right)
,
\ea
}
where $x_1, t_1, t_s, y_s, y_2, ..., s_E$ are
 the time instances of transitions multiplied by $k$.
See Appendix C.
(In (43) and (44) of Ref.\cite{WangZhangChen2016}
$x_1$ and $s_E$ should have minus signs.)
We plot  $|\alpha_k|^2$ and $|\beta_k|^2$
as functions of $k$ in Fig.\ref{alphakbetak}.
They satisfy the relation
\be \label{bogolyubov}
|\alpha_k|^2-|\beta_k|^2=1,
\ee
which is implied by applying the  Wronskian (\ref{wronsk}).
The relation (\ref{bogolyubov})
can be checked to be satisfied by (\ref{e1}) and (\ref{e2})
to each order of powers of $k$.
The positive frequency mode in (\ref{upresent}) is taken as
the vacuum mode at the present stage
\be \label{uvacpres}
v_{k}(\tau )= \sqrt{\frac{\pi}{2}}\sqrt{\frac{s}{2k}}
          e^{i\pi \gamma/2} H^{(2)}_{-\gamma-\frac{1}{2}}(s),
\ee
such  that
$v_{k}(\tau ) \rightarrow \frac{1}{\sqrt{2k}} e^{-ik(\tau-\tau_a)} $
as  $k\rightarrow\infty$.
Then the mode solution (\ref{upresent}) is written   as
\be \label{upresent2}
u_k(\tau ) = \alpha_k v_k(\tau) +\beta_k  v_k^*(\tau) .
\ee
Starting from the vacuum fluctuation  during inflation
with only positive-frequency modes (\ref{u}),
RGW has  evolved into a mixture of the positive and negative frequency modes
as in Eq.(\ref{upresent2}).
The field operator $h_{ij}$ in Eq.(\ref{Fourier}) in the present stage,
for  each  $\bf k$ and each $s$ polarization,
  is proportional to
\ba
&& \left[ a^s_{\bf k}   h^s_k(\tau)e^{i\bf{k}\cdot\bf{x}}
    +a^{s\dagger}_{\bf k}h^{s*}_k(\tau)e^{-i\bf{k}\cdot\bf{x}}\right]
       \nn \\
&& = \frac{A}{a(\tau)}
 \left[  a_{\bf k}    \alpha_k v_k  e^{i\bf{k}\cdot\bf{x}}
       + a^{\dagger}_{\bf k} \beta_k^*  v_k  e^{-i\bf{k}\cdot\bf{x}}
        +  a_{\bf k}\beta_k  v_k^*  e^{i\bf{k}\cdot\bf{x}}
        +  a^{\dagger}_{\bf k} \alpha_k^* v_k^*  e^{-i\bf{k}\cdot\bf{x}}
       \right] \nn \\
&& = \frac{A}{a(\tau)}
   \l[ A_{\bf k}   v_k  e^{i\bf{k}\cdot\bf{x}}
      +  A_{\bf k}^{\dagger}     v_k^*  e^{-i\bf{k}\cdot\bf{x}} \r] . \nn
\ea
The above terms will appear in the summation over $\bf k$,
so one can change the sign of the wavevector $\bf k$ in the $\beta_k$ terms.
Thus
\be\label{FourierPresent}
h_{ij}  ({\bf x},\tau)
=\int\frac{d^3k}{(2\pi)^{3/2}}
    \sum_{s={+,\times}} {\mathop \epsilon  \limits^s}_{ij}(k)
    \frac{A}{a(\tau)}
    \l[ A_{\bf k}   v_k(\tau)  e^{i\bf{k}\cdot\bf{x}}
      +  A_{\bf k}^{\dagger}     v_k^*(\tau)  e^{-i\bf{k}\cdot\bf{x}} \r]       ,
\ee
where
\be\label{Akak}
 A_{\bf k} \equiv \alpha_k  a_{\bf k}  + \beta^*_k a^\dagger_{-\bf k}
\ee
is  interpreted as the annihilation operator of gravitons of $\bf k$
for the present   stage.
(In the expression of $A_{\bf k}$ in Ref.\cite{WangZhangChen2016}
where the subscript should have a minus sign,
$a^\dagger_{\bf k}\rightarrow a^\dagger_{-\bf k}$.)
The number density of gravitons  in $k$-mode in the present  stage is
\be
N_{\bf k } = \langle 0|  A_{\bf k}^\dagger   A_{\bf k} |0\rangle = |\beta_k|^2,
\ee
which has been generated during the expansion.

By   Eqs.(\ref{bogolyubov}) (\ref{upresent2}),
we write the  power spectrum (\ref{spectrum}),
the spectral energy density (\ref{rhok})
and pressure (\ref{pressFPth})  at the present time
\ba  \label{unregulspectrum}
\Delta_t^2(k,\tau_H)
& = & \frac{A ^2 k^3}{\pi^2 a^2 }
 \left( |v_k|^2 + 2Re (\alpha_k \beta_k^* v_k^2 )
          + 2|\beta_k|^2 |v_k|^2 \right) ,
\ea
\be  \label{rhokacc}
\rho_k(\tau_H) = \frac{k^3}{\pi^2a^2}
  \Bigg(  \left| ( \frac{v_k }{a} ) '  \right|^2
        +   2Re[ \alpha_k \beta_k^*   ( \frac{v_k }{a} ) ^{\prime\,  2}]
        + 2|\beta_k|^2   \left| ( \frac{v_k }{a} ) '  \right|^2 \Bigg),
\ee
\be\label{prpres}
p_k(\tau_H) =\frac{k^5}{3\pi^2a^4}
 \left(|  v_k |^2
  +   2Re ( \alpha_k \beta_k^*    v_k ^{2} )
       + 2|\beta_k|^2   \left| v_k \right|^2  \right),
\ee
all of them contain three terms.
Consider Eq.(\ref{unregulspectrum}) as  example,
the first term $|v_k |^2 $ is the present vacuum contribution.
$ 2|\beta_k|^2 |v_k|^2$ is the graviton contribution.
$2Re (\alpha_k \beta_k^* v_k^2 )$ is the interference
between  the positive and negative frequency modes.
(It  was also called as
the vacuum-graviton coupling in Paper I.)
These interpretations also apply to (\ref{rhokacc}) and (\ref{prpres}).
The interference arises inevitably as soon as
the negative frequency modes $\beta_k v_k^*$ are developed
in the reheating stage after inflation
and in the subsequent stages.
Hence, the particle production in the expanding RW  spacetime
is always accompanied by the interference.
In this sense, the interference  is a prediction of  quantum field theory
in curved spacetime.

\begin{figure}
\centering
\includegraphics[width=0.7\linewidth]{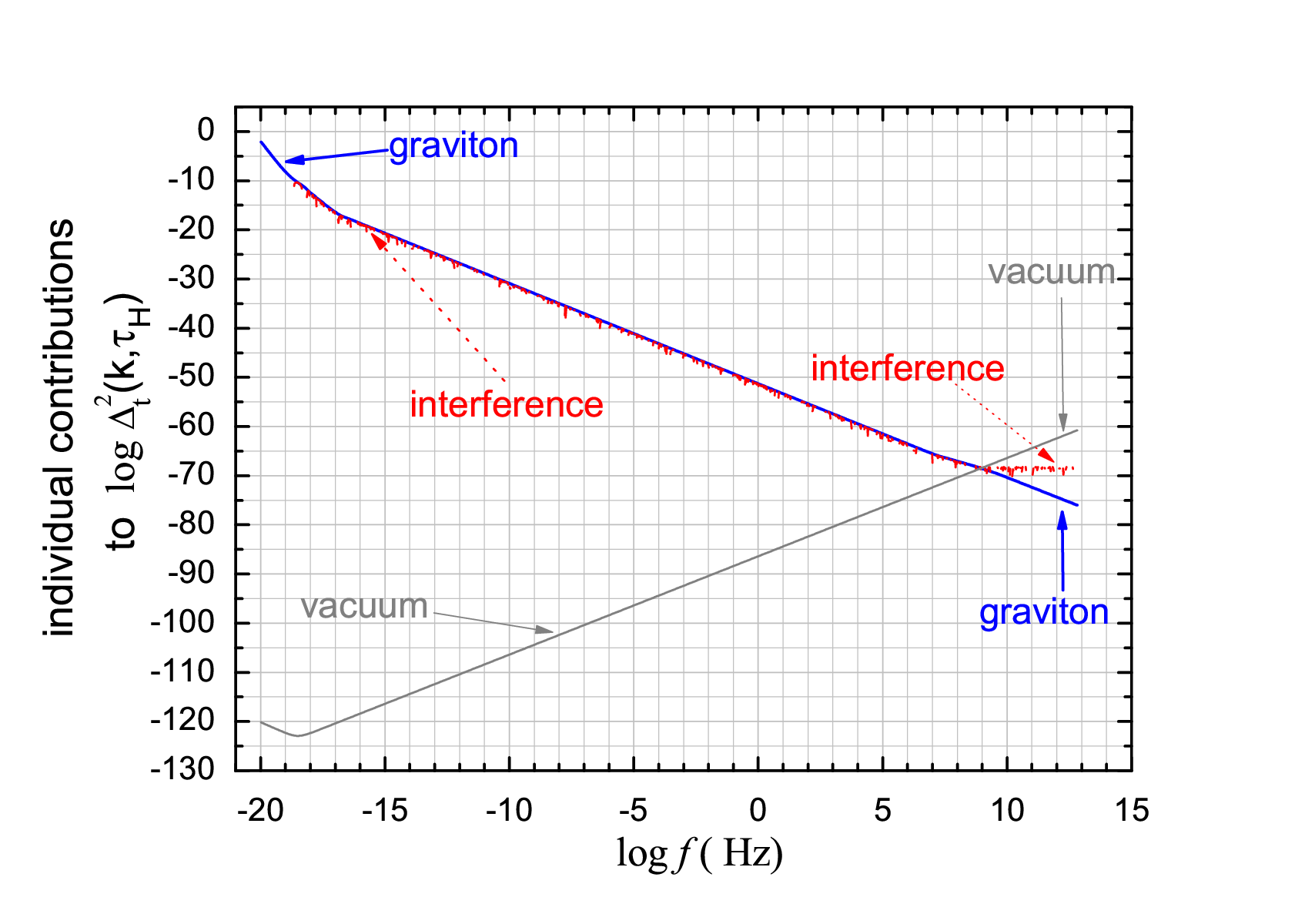}
\caption{  The individual contributions to  unregularized power spectrum
    by the vacuum, interference, and graviton.
    The interference has negative values which are not shown in log plot.
    }
     \label{sp3terms2}
\end{figure}

The contributions of the three terms
vary in different frequency ranges and are plotted in Fig.\ref{sp3terms2}
for the unregularized power spectrum.
Over the  range $f\leq 10^{9}$Hz,
the graviton term is dominant,
the oscillatory interference term is comparable,
and the vacuum term is negligibly small.
For   $f \geq 10^{9}Hz$,
    the vacuum dominates and is $\propto k^2$ quadratic divergent,
    the interference is $\propto k^0$ logarithmic divergent,
    the graviton is  $\propto k^{-2}$  UV convergent.

\begin{figure}
\centering
\includegraphics[width=0.7\linewidth]{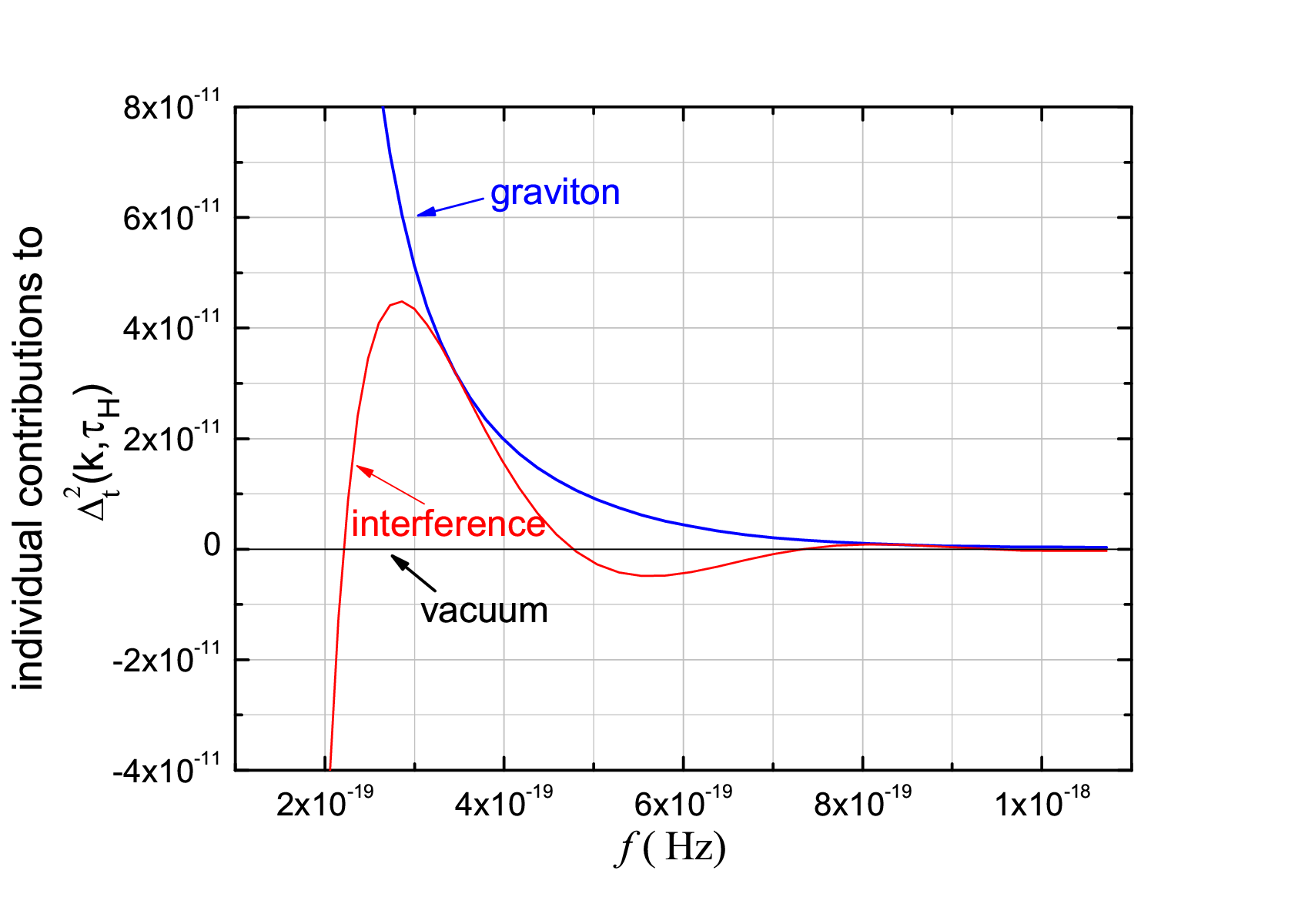}
\caption{ The enlarged portion of Fig. \ref{sp3terms2}
    at low frequency end $f \lesssim 10^{-17}$Hz,
      corresponding to large angular CMB anisotropies and polarization.
    The graviton dominates,
    the interference is oscillatory,
    the vacuum is negligibly small.
        }
     \label{sp3termsLow2}
\end{figure}

Fig.\ref{sp3termsLow2} is an enlarged portion of Fig.\ref{sp3terms2}
for $f \lesssim 10^{-17}$Hz.
This portion  corresponds to
the range probed by CMB anisotropies
\cite{WangZhangChen2016,XiaZhang20089A,XiaZhang20089B}.

\begin{figure}
\centering
\includegraphics[width=0.9\linewidth]{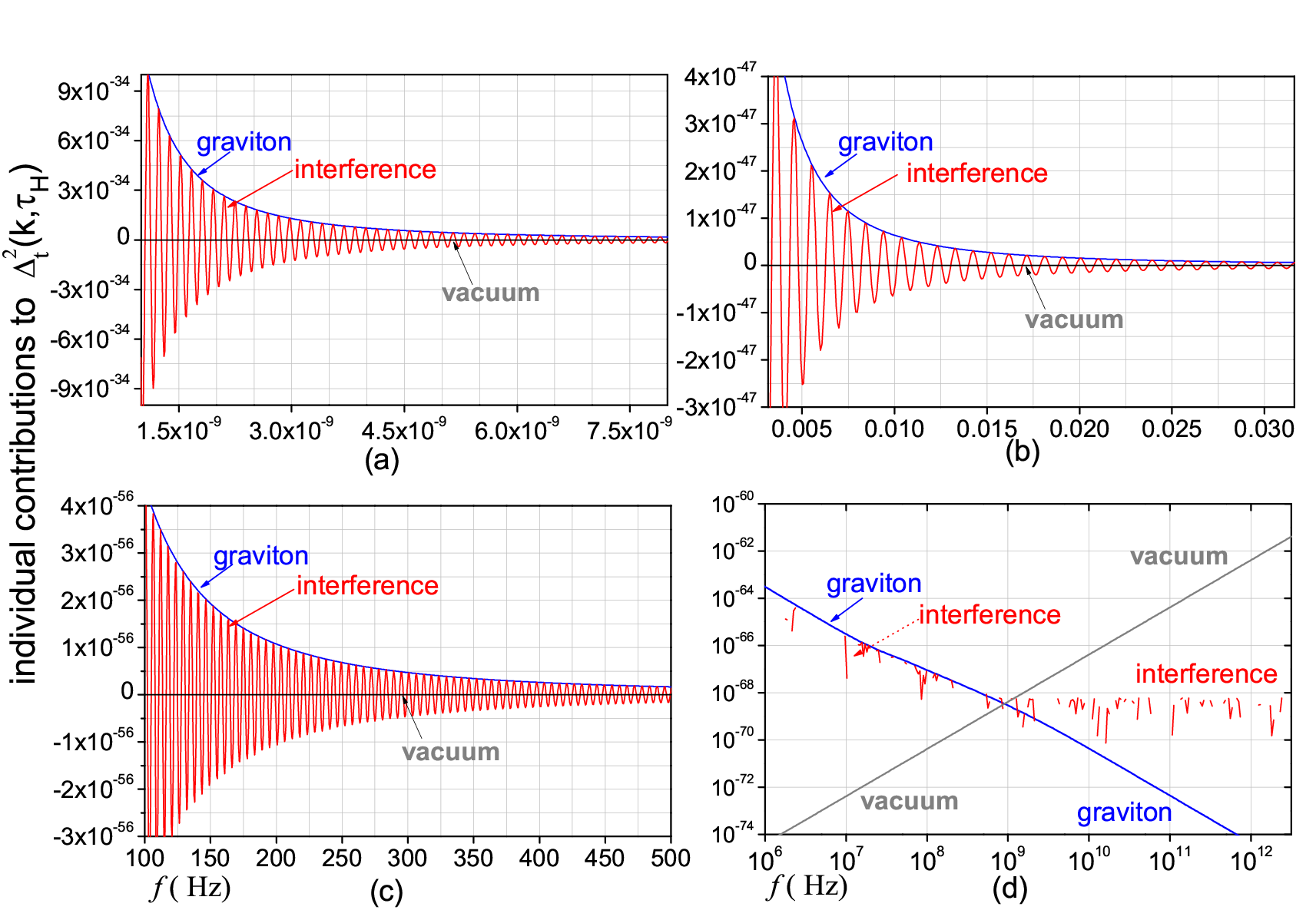}
\caption{  (a): Around $f \sim  10^{-9}$Hz,
         for PTA.
(b): Around $f \sim 10^{-2}Hz$,
         for LISA.
(c):  Around $f \sim 10^{2}Hz$,
         for LIGO.
(d): Around  $f \sim 10^{9}Hz$, for Gauss-beam detector.
        }
     \label{reg4freq}
\end{figure}

Fig.\ref{reg4freq} (a) shows the  range around $f \sim  10^{-9}$Hz
corresponding to that of PTA detectors.
In this range the characteristic amplitude
$h(f,\tau_H)\equiv \sqrt{\Delta_t^2(k,\tau_H)}\sim 10^{-17}$
is high  and may be possibly detected by PTA detectors,
such as PPTA, EPTA, SKA, NANOGrav, FAST, etc \cite{SKA-Smits2009A,SKA-Smits2009B,
EPTA-Haasteren2011,Nan2011,NANOGrav-Demorest2013,Tong2014}.
The oscillation frequency of interference is $\sim 8 \times f \sim 10^{-8}$Hz and
the oscillatory amplitude is $\sim 10^{-33}$ decreasing at large $f$.
This unique oscillatory feature existing
will be helpful to distinguish the RGW signal from the GW foreground.

Fig.\ref{reg4freq} (b) shows   the  range around $ f \sim  10^{-2}$Hz
corresponding to LISA \cite{LISAwebA,LISAwebB}.
The characteristic amplitude $h(f,\tau_H) \sim 10^{-23}$,
the oscillation frequency  $\sim 10 \times f \sim 10^{-1}$Hz.

Fig.\ref{reg4freq} (c) shows  the   range around $f \sim  10^{2}$Hz
corresponding to LIGO \cite{LIGOweb},
the oscillation frequency  $\sim 18 \times f \sim 10^{3}$Hz.

Fig.\ref{reg4freq} (d) shows the high frequency portion
around $f \sim  10^{9}Hz$.
This may be  explored
high-frequency Gaussian beam detectors \cite{Li2003,TongZhangGaussian}.

From Figs.\ref{reg4freq} (a),(b),(c)
one sees that
the oscillation amplitude of interference is comparable to that of the vacuum,
the latter actually forms the upper envelope of the oscillatory interference.
The oscillation amplitude decreases with
and the oscillation frequency  increases  with the spectrum frequency $f$.
However, the details of interference depend on the regularization scheme,
and its extent may be changed in other schemes.
Future detections of RGW  will have chance to probe these.


\section{Regularization at the present time}

As possible alternative to that in Section 5,
we try to perform regularization at the present time.
The three terms in  the power spectrum (\ref{unregulspectrum})
need to be regularized.
As before, the vacuum term is subtracted by  $ |v_k^{(2)}|^2$
which is the 2nd adiabatic order counter term for the present stage
(see  (\ref{counter2ndacc}) in  Appendix A).
The interference term  $Re (\alpha_k \beta_k^* v_k ^2)$
is subtracted by $(v_k ^{(0)})^2$,
which is the 0th order counter term given in (\ref{counter0th}).
The graviton  term $|\beta_k|^2  |v_k |^2 \propto k^{-5}$
 is already UV convergent and needs no regularization.
So the  power spectrum is regularized at the present time
as  the following
\ba  \label{regspectr}
\Delta_t^2(k,\tau_H)_{reg}
 & = &   \frac{A ^2 k^3}{\pi^2 a^2(\tau_H)}
 \Bigg[ \left(|v_k|^2 -(\frac{1}{2k}
       +\frac{\gamma (\gamma +1)}{4k^3  (\tau_H-\tau_a)^2}) \right) \nn \\
&&     + 2Re \left(\alpha_k \beta_k^*
          ( v_k^2 - \frac{e^{-2ik(\tau_H -\tau_a)}}{ 2k} ) \right)
       + 2|\beta_k|^2 |v_k|^2 \Bigg].
\ea
The spectral energy density (\ref{rhokacc})
and pressure (\ref{prpres}) are regularized up to the 4th adiabatic order
\ba\label{ereg}
\rho_k(\tau_H)_{reg} & = & \frac{k^3}{\pi^2a^2}
  \Bigg|  \left| ( \frac{v_k }{a} ) '  \right|^2
    -  \Big | ( \frac{v_k^{(4)}  }{a} ) ' \Big|^2
     +  2Re \Big[ \alpha_k \beta_k^*
 \Big( (\frac{v_k}{a} )^{\prime\, 2} - (\frac{v_k^{(2)} }{a}) ^{\prime\,  2 }
              \Big)      \Big]     \nn \\
&&   + 2|\beta_k|^2  \Big( \left| ( \frac{v_k }{a} ) '  \right|^2
  -   | ( \frac{v_k^{(0)}}{a} ) ' |^2  \Big) \Bigg|,
\ea
\ba \label{prega}
p_k(\tau_H)_{re} & = & \frac{k^5}{3\pi^2a^4}
  \Big||v_k|^2 - |v_k^{(4)}|^2
  +   2Re \l[ \alpha_k \beta_k^*   \big(v_k ^{2}-  ( v^{(2) }_k  )^2 \big) \r] \nn \\
  & &    + 2|\beta_k|^2  ( |v_k  |^2 - |v_k^{(0)}|^2 )  \Big | ,
\ea
where  six counter terms
are listed  in (\ref{vct4or})--(\ref{vcount0th}) in Appendix A,
giving the explicit formulae
\ba  \label{rhokacc4k}
\rho_k(\tau_H)_{reg} & = & \frac{k^3}{\pi^2a^2}
 \Big|
   | ( \frac{v_k }{a} ) ' |^2 -
  \frac{k}{a^2}
  \Big(\frac{1}{2}
   +\frac{\gamma(\gamma-1)}{4k^2 \tau^2}
   +\frac{3(\gamma-2)(\gamma-1)\gamma(\gamma+1)}{16k^4  \tau^4} \Big)  \nn \\
&  & +   2Re \Big( \alpha_k \beta_k^*
   \Big[ \l( \frac{v_k }{a} \r) ^{\prime\,  2}   - \frac{k  e^{-2ik\tau}}{2  a^2}
    \Big( -1 + i\frac{\gamma(\gamma-1)}{k\tau} \nn \\
 && ~  +  \frac{\gamma(\gamma-1)(\gamma^2-\gamma-1)}{2k^2\tau^2}  \Big)  \Big] \Big)
   + 2|\beta_k|^2 \Big( | ( \frac{v_k }{a} ) ' |^2
         - \frac{k}{ 2a^2 } \Big)  \Big| ,
\ea
\ba\label{pressRegTauHk}
p_k(\tau_H)_{re} &  = & \frac{k^5}{3\pi^2a^4}
 \Big| |v_k|^2 - \frac{1}{ 2k}
\left( 1 + \frac{\gamma(\gamma+1)}{2k^2\tau^2}
      + \frac{3(\gamma+2)(\gamma+1)\gamma(\gamma-1)}{8 k^4 \tau^4} \right)
         \nn \\
& &  +  2Re \left( \alpha_k \beta_k^*   \Big[ v_k ^{2}- \frac{e^{-2ik\tau}}{2k}
   \Big( 1 - i\frac{\gamma(\gamma+1)}{k\tau}
     -\frac{\gamma(\gamma+1)(\gamma^2+\gamma-1)}{2 k^2 \tau^2} \Big) \Big] \right)
                \nn \\
&&   + 2|\beta_k|^2     \Big(|v_k|^2 - \frac{1}{2k} \Big)  \Big|  .
\ea
In the above, $\tau$ stands for  $(\tau_H-\tau_a)$ to avoid clumsy notation.

\begin{figure}
\centering
\includegraphics[width=0.9\linewidth]{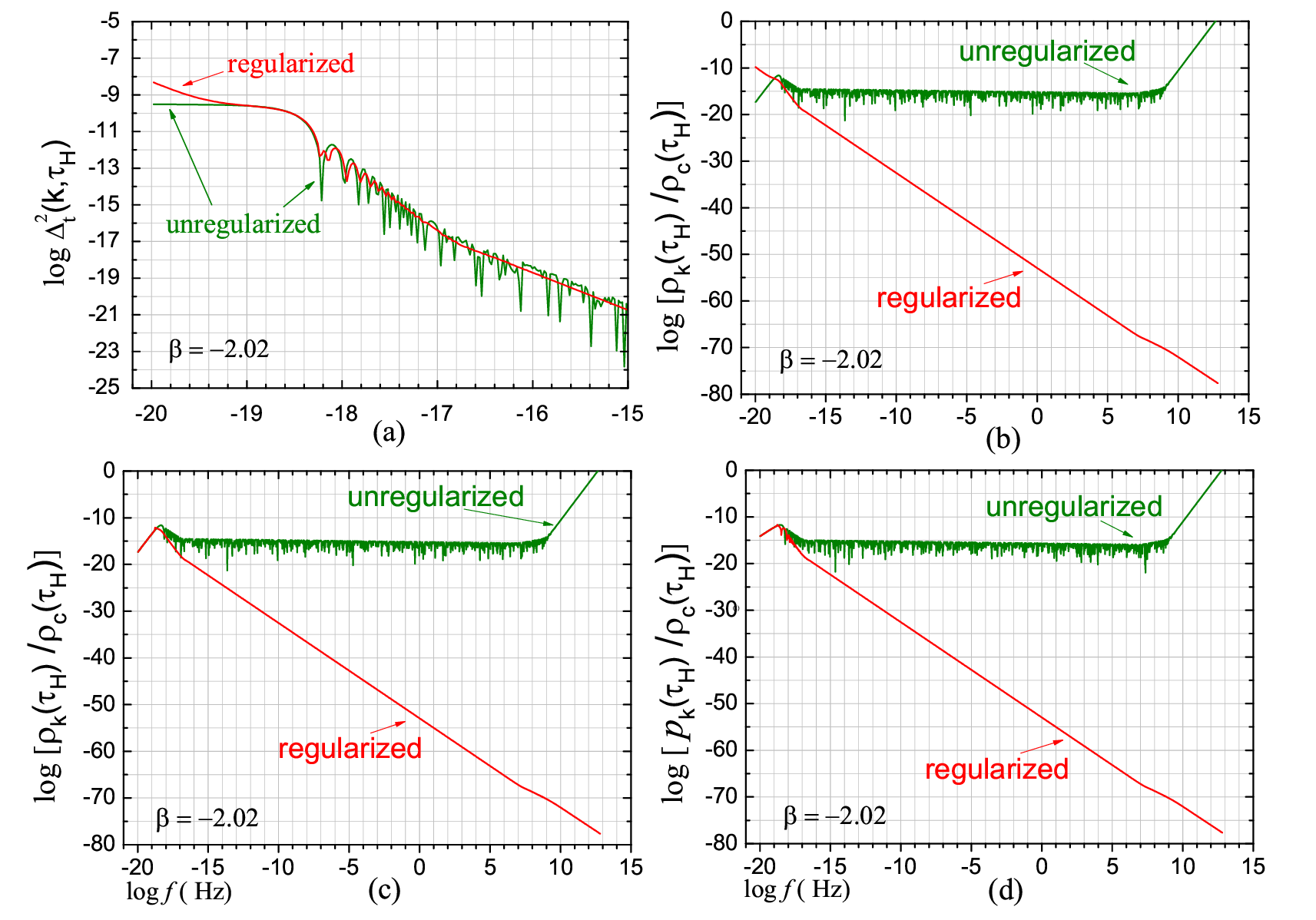}
\caption{
(a) $\Delta_t^2$ regularized  at $\tau_H$ for all $k$
 has distortions at low frequency end.
(b) $\rho_k$ regularized   at $\tau_H$ for all $k$
       is drastically  distorted.
(c) (d) $\rho_k$ and $p_k$
    regularized  at $\tau_H$ for $k> \frac{1}{|\tau_H-\tau_a|}$
    are  also drastically distorted.
    }
     \label{4regtH19}
\end{figure}

If the formulae (\ref{regspectr}) (\ref{rhokacc4k}) (\ref{pressRegTauHk}) are applied
for all $k$-modes,
the resulting power spectrum shown in Fig.\ref{4regtH19} (a)
is uplifted at $k\sim 0$,
the interference oscillations are suppressed.
The  resulting $\rho_k(\tau_H)_{reg} $ is
totally distorted as  shown in Fig.\ref{4regtH19} (b).
The pressure has a similar situation.
Thus, the all-$k$ regularization at the present time is unsuccessful.

If  (\ref{regspectr})(\ref{rhokacc4k})  (\ref{pressRegTauHk})
are applied to the modes inside the present horizon $k> \frac{1}{|\tau_H-\tau_a|}$,
corresponding to  $f> \frac{H_0}{2\pi}  \sim    10^{-19}$Hz.
Again, $\rho_k(\tau_H)_{reg} $
is  drastically distorted, as shown in Fig.\ref{4regtH19} (c).
This scheme  is unsuccessful either.

\begin{figure}
\centering
\includegraphics[width=0.9\linewidth]{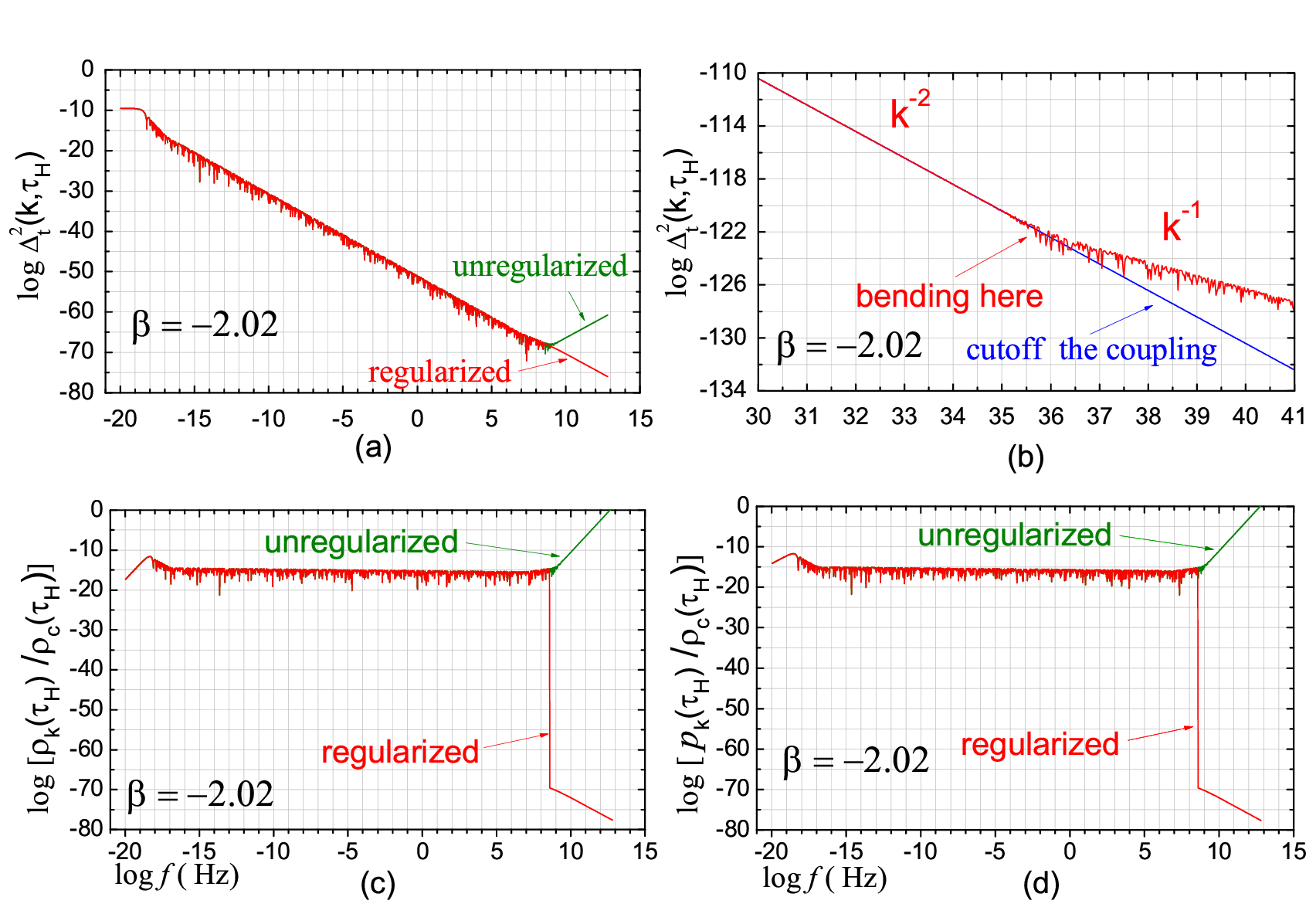}
\caption{
(a)  $\Delta_t^2(k,\tau_H)_{reg}$  regularized  at $\tau_H$
      for $f>10^9$Hz.
(b)   The high frequency end of (a)
       has a bending at $f\sim 10^{35}$Hz.
(c) (d) $\rho_k(\tau_H)_{reg}$ and $p_k(\tau_H)_{reg}$
           regularized  at $\tau_H$  for $f>10^9$Hz.
         }
     \label{regtH4inH}
\end{figure}

An inspection tells that
all three  unregularized   spectra
  rise up at  $f \gtrsim 10^9$Hz
which corresponds to  $k > 1/|\tau_1|$ by the relation (\ref{fk}).
So,
the present $k$-modes that carry UV divergences
are identified as the same modes that were carrying UV divergences during inflation.
Therefore,
we try to  apply the formulae
(\ref{regspectr}) (\ref{rhokacc4k}) (\ref{pressRegTauHk})
only to the high $k$-modes with
$f > 10^{9} ~ {\rm Hz}$,
and hold the low $k$  intact.
The resulting power spectrum shown in Fig.\ref{regtH4inH} (a)
has no distortion for   $f<  10^{9}$Hz,
similar to Fig.\ref{specInfToAcc2}.
However, at high frequency as shown in Fig.\ref{regtH4inH} (b),
it behaves  as $\propto k^{-2}$ in the range $(10^{9} \sim 10^{35})$Hz,
and as  $\propto k^{-1}$ for  $f> 10^{35}$Hz,
giving rise to  a bending.
This differs from  an unbending result $\propto k^{-2}$
in Paper I where a cutoff was made on the interference term.
The spectral energy density and pressure
shown in Fig.\ref{regtH4inH} (c) (d)
have no distortion for  $f< 10^{9}$Hz,
similar to  Fig.\ref{rhopressure4},
but  have an abrupt drop at $f \simeq  10^{9}$Hz.
Thus, this scheme
yields the high-frequency irregularities, which seem artificial.
The actual spectra at high frequency
may eventually  be explored
by high-frequency Gaussian beam detectors \cite{Li2003,TongZhangGaussian}.

\section{Conclusion and Discussion}

The power spectrum of RGW
contains UV divergences  up to 2nd adiabatic  order,
and the energy density and pressure  of RGW
 contain UV divergences up to 4th order,
the scheme of all-$k$  regularization removes UV divergences,
but  also brings about  distortions at low frequencies.
More severely,
as our  analysis has revealed,
the  spectral energy density and pressure
are even changed to IR divergent
under the all-$k$  regularization.

To avoid these low-frequency distortions,
we have proposed the scheme of inside-horizon regularization
at a fixed instance during inflation.
This is motivated by the fact that
the UV divergences are due to the short-wavelength modes inside the horizon,
whereas the long-wavelength modes are not responsible.
We regularize only the inside-horizon modes and keep the outside-horizon  modes intact.
Consequently,
the regularized spectra all become UV convergent,
and  simultaneously are free of IR distortions
for a whole  range of $f\lesssim 10^{9}$Hz,
as shown
in Figs.\ref{RiseupRegInH1-1000},  \ref{t1pwreg2198},  \ref{infend4all}.
With these spectra  as the initial condition,
we let them evolve, and obtain the present spectra
in Fig.\ref{specInfToAcc2} and  Fig.\ref{rhopressure4},
which all remain  UV convergent and free of IR distortion.
For  de Sitter inflation,
the scheme yields the  spectra,
which are non-vanishing and free of IR distortion,
as shown in Fig.\ref{specInfToAccbeta2} and  Fig.\ref{rhopressure4},
thus, overcoming  the difficulty with
the all-$k$  regularization
that would inevitably lead to vanishing spectra.
It is legitimate to adopt
the inside-horizon scheme of regularization for RGW,
because at the level of the linearized  Einstein equation,
$k$-modes of RGW are linearly independent.
The inside-horizon scheme can apply also to other linear fields
such as the scalar curvature perturbation and
the gauge-invariant perturbed scalar field during inflation.

Three  schemes of regularization at the present time
have been explored and are found  to give some irregularities.
Thus, the inside-horizon scheme at the end of inflation
is  preferable to these three,
since it yields the UV-convergent spectra from start
which are free of low-frequency distortion.
Nevertheless,
even our inside-horizon scheme presented in this paper
may be not  the final prescription,
other possible valid schemes can also  exist,
because generally  there is no unique adiabatic regularization
from the  perspective of renormalization.

We have also  analyzed the structure of spectra of RGW as quantum field
at the present stage,
consisting of the graviton, the interference, and  the vacuum contributions.
The former two contributions are dominant in the   range $f\lesssim 10^{9}$Hz,
which covers almost all the observational band of detectors.
As a prominent feature,
the spectra contain quick  oscillations
in the frequency domain, as plotted in Figs.\ref{sp3termsLow2} and \ref{reg4freq}.
This  oscillatory pattern is  due to
interference of positive and negative frequency modes of RGW.
Since the interference depends on the regularization schemes,
and no final conclusive scheme has been arrived so far,
we shall leave this to be probed by future RGW detections.

\

\textbf{Acknowledgements}

Y. Zhang is supported by
NSFC Grant No. 11421303, 11675165, 11633001
SRFDP, and CAS, the Strategic Priority Research Program
``The Emergence of Cosmological Structures"
of the Chinese Academy of Sciences, Grant No. XDB09000000.

\

\appendix
\numberwithin{equation}{section}

\
\\
{\bf \Large Appendix}

\section{The adiabatic counter terms up to 4th order }

In this appendix,
we calculate the adiabatic counter terms
and list explicitly their analytical expressions
that have been used in the context.
See also Refs. \cite{Chakraborty1973,
ParkerFulling1974,Bunch1980}
for a scalar field.
For RGW in a RW spacetime,
the $n$-th adiabatic mode for an integer $n \ge 0$ is
\be\label{nthadi}
v_k^{(n)}(\tau)= \frac{1}{\sqrt{2W_k^{(n)}(\tau)}}
          \exp\left[-i\int_{\tau_0}^{\tau}W_k^{(n)}(\tau')d\tau'\right]  ,
\ee
where
\be \label{wn}
W_k^{(n)}=\sqrt{k^2-\frac{a''}{a}
-\frac{1}{2}\left[\frac{W_k^{(n-2)\prime\prime}}{W_k^{(n-2)}}
-\frac{3}{2}\left(\frac{W_k^{(n-2)\prime}}{W_k^{(n-2)}}\right)^2\right]}.
\ee
To the 0th order,   $W_k^{(0)}=k$,
$v_k^{(0)}=\frac{1}{\sqrt 2k}e^{-ik\tau}$,
which corresponds to the mode in Minkowski spacetime.
To the 2th adiabatic order  $W_k^{(2)}  =  \sqrt{k^2- \frac{a''}{a}} $,
the 4th adiabatic order
\ba
W_k^{(4)} & =  & k \sqrt{1-\frac{a''}{k^2 a}-\frac{1}{4k^4 a^2}
(a''^2-aa''''+2a'a'''-2\frac{a\,'^2a''}{a} ) } .
\ea
By the construction,
the 4th order adiabatic  mode $v_k^{(4)}$ satisfies
the wave equation (\ref{evolution}) to 4th order,
and respects the covariant conservation to 4th order.
Similar  for the 2nd order adiabatic mode $v_k^{(2)}$.

For  the accelerating stage $a(\tau) \propto |\tau|^{-\gamma}$,
\ba
W_k^{(4)}  &= & k \sqrt{1-\frac{\gamma(\gamma+1)}{k^2 \tau^2 }
      +\frac{6 \gamma (\gamma+1)}{4k^4 \tau^4}  } \nn \\
& \simeq &     k \left(1 -  \frac{\gamma(\gamma+1)}{2  k^2 \tau^2}
     -\frac{\gamma (\gamma-2)(\gamma+1)(\gamma +3)}{8k^4\tau^4}
       \right) .
\ea
Substituting this into (\ref{nthadi})
yields the 4th adiabatic  mode
\ba
v_k^{(4)}(\tau)
& \simeq  &\frac{e^{-ik\tau}}{\sqrt{2k}}
  \Big( 1 -i   \frac{\gamma(\gamma+1)}{2 k \tau}
    -\frac{(\gamma+2)(\gamma+1)\gamma(\gamma-1)}{8k^2\tau^2}
              \nn \\
 && +i \frac{(\gamma-2)(\gamma-1)\gamma(\gamma+1)(\gamma+2)(\gamma+3)}{48 k^3 \tau^3}
    \Big) ,
    \label{vadto4}
\ea
and the 2nd order adiabatic counter mode follows
\be
v_k^{(2)}(\tau)  =
  \frac{e^{-ik\tau}}{\sqrt{2k}}
  \Big( 1 -i   \frac{\gamma(\gamma+1)}{2 k \tau}
    -\frac{(\gamma+2)(\gamma+1)\gamma(\gamma-1)}{8k^2\tau^2} \Big).
\ee
From  these counter modes,
one obtains the   adiabatic counter terms
that are used in the context.
The  time derivatives are
\be\label{vct4or}
 \big| ( \frac{v_k^{(4)}(\tau) }{a} ) ' \big|^2
   =\frac{1}{a^2}
   \left[\frac{k}{2}
   +\frac{\gamma(\gamma-1)}{4k\tau^2}
   +\frac{3(\gamma-2)(\gamma-1)\gamma(\gamma+1)}{16k^3\tau^4}\right],
\ee
\be\label{vsq2ex}
(\frac{v_k^{(2)} }{a}) ^{\prime\,  2 }
 =    \frac{ k e^{-2ik\tau}}{ 2 a^2 }
   \l( -1 + i\frac{\gamma(\gamma-1)}{k\tau}
  +  \frac{\gamma(\gamma-1)(\gamma^2-\gamma-1)}{2k^2\tau^2} \r),
\ee
\be\label{v0t}
 \big| ( \frac{v_k^{(0)}(\tau) }{a} ) ' \big|^2
   = \frac{1}{a^2}    \frac{k}{2} ,
\ee
for the  energy density.
\ba\label{vcount4sq}
|v_k^{(4)}(\tau)|^2
     &   =  &   \frac{1}{ 2k}
\Big[ 1 + \frac{\gamma(\gamma+1)}{2k^2\tau^2}
      + \frac{3(\gamma+2)(\gamma+1)\gamma(\gamma-1)}{8 k^4 \tau^4} \Big] ,
\ea
\be\label{vcount2}
\big( v^{(2) }_k (\tau) \big)^2
= \frac{e^{-2ik\tau}}{2k}
  \Big( 1 - i\frac{\gamma(\gamma+1)}{k\tau}
      -\frac{\gamma(\gamma+1)(\gamma^2+\gamma-1)}{2 k^2 \tau^2} \Big)  ,
\ee
\be \label{vcount0th}
|v_k^{(0)}(\tau)|^2    =  \frac{1}{ 2k} ,
\ee
for the pressure.
And
\be \label{counter2ndacc}
|v_k^{(2)}(\tau)|^2
        =     \frac{1}{ 2k}
\Big( 1 + \frac{\gamma(\gamma+1)}{2k^2\tau^2}    \Big),
\ee
\be \label{counter0th}
(v_k^{(0)}(\tau))^2 =  \frac{e^{-2ik \tau }}{ 2k} ,
\ee
for the power spectrum.

For inflation $a(\tau) \propto |\tau|^{\beta+1}$,
just replacing $\gamma \rightarrow -\beta-1$ in the above,
one obtains
the following adiabatic counter terms
\be\label{ad4v}
 \big| ( \frac{v_k^{(4)}(\tau) }{a} ) ' \big|^2
   =\frac{1}{a^2}\left[\frac{k}{2}+\frac{(\beta+1)(\beta+2)}{4k\tau^2}
 +\frac{3\beta(\beta+1)(\beta+2)(\beta+3)}{16k^3\tau^4}\right]
\ee
for the energy density,
\be\label{4thorderterm2}
 \big| v_k^{(4)}(\tau)    \big|^2
  = \frac{1}{2 k}  + \frac{\beta(\beta+1) }{4 k^3 \tau^2}
    + \frac{3(\beta -1 ) \beta (\beta+1) (\beta+2)}{16 k^5 \tau^4}
\ee
for the pressure, and
\be\label{v2ex}
 \left| v_k^{(2)}(\tau)    \right|^2
  = \frac{1}{2 k}  + \frac{\beta(\beta+1) }{4 k^3 \tau^2}
\ee
for the power spectrum.

\section{The solution mode at high frequency  }

It is revealing to expand the RGW mode solutions $v_k$  at high frequency
in terms of  powers of $k$,
in comparing with the adiabatic counter terms.
For  the present accelerating stage,
the vacuum mode of RGW $v_k$ in  (\ref{uvacpres})  at high frequencies
can be expanded as powers of $k$ as the following
{\allowdisplaybreaks
\ba\label{vex4}
v_k(\tau)&=&\frac{e^{-ik\tau}}{\sqrt{2k}}
\Big(1-i\frac{\gamma(\gamma+1)}{2k\tau}
-\frac{(\gamma+2)(\gamma+1)\gamma(\gamma-1)}{8k^2\tau^2}\nn\\
&&+i\frac{(\gamma+3)(\gamma+2)(\gamma+1)\gamma(\gamma-1)(\gamma-2)}{48k^3\tau^3} \nn \\
&& + \frac{(\gamma+4)(\gamma+3)(\gamma+2)(\gamma+1)\gamma(\gamma-1)(\gamma-2)(\gamma-3)}{384 k^4\tau^4}
             +,,,  \Big) ,
  \nn \\
\ea
\ba\label{vsqaccex}
|v_k(\tau)|^2
     &   =  &   \frac{1}{ 2k}
\Big[ 1 + \frac{\gamma(\gamma+1)}{2k^2\tau^2}
      + \frac{3(\gamma+2)(\gamma+1)\gamma(\gamma-1)}{8 k^4 \tau^4} \nn \\
    & & + \frac{5(\gamma+3)(\gamma+2)(\gamma+1)\gamma(\gamma-1)(\gamma-2)}{16 k^6 \tau^6}
         +,,,    \Big] ,
\ea
\ba\label{v2expd}
v^2_k(\tau) & = & \frac{e^{-2ik\tau}}{2k}
  \Big( 1 - i\frac{\gamma(\gamma+1)}{k\tau}
      -\frac{\gamma(\gamma+1)(\gamma^2+\gamma-1)}{2 k^2 \tau^2} \nn \\
 &&  + i \frac{(\gamma-1)\gamma(\gamma+1)(\gamma+2)(2\gamma^2+2\gamma-3)}{12 k^3 \tau^3}
    +,,,    \Big)      .
\ea
}
The derivatives are
\ba\label{v2pri}
\l( \frac{v_k }{a} \r) ' &  =&
\frac{e^{-ik\tau}}{a  \sqrt{2k}}
\Big(-i k -\frac{\gamma(\gamma-1)}{2 \tau}
 +i \frac{(\gamma+1)\gamma(\gamma-1)(\gamma-2)}{8k\tau^2} \nn \\
& & + \frac{(\gamma+2)(\gamma+1)\gamma(\gamma-1)(\gamma-2)(\gamma-3)}{48k^2\tau^3} \nn \\
& & -i  \frac{(\gamma+3)(\gamma+2)(\gamma+1)\gamma(\gamma-1)
  (\gamma-2)(\gamma-3)(\gamma-4)}{384 k^3\tau^4} +,,,  \Big ) ,
  \nn
\ea
\ba \label{vsq4}
 \Big| ( \frac{v_k}{a} ) '  \Big|^2
 &=&
 \frac{1}{a^2}\left[\frac{k}{2}
 +\frac{\gamma(\gamma-1)}{4k\tau^2}
 +\frac{3(\gamma-2)(\gamma-1)\gamma(\gamma+1)}{16k^3\tau^4}\right. \nonumber  \\
&& \left.
  +\frac{5(\gamma-3)(\gamma-2)(\gamma-1)\gamma(\gamma+1)(\gamma+2)}{32k^5\tau^6}
     +,,, \right] ,
\ea
\ba \label{v2prisqr}
\left( ( \frac{v_k }{a} ) ^{'} \right)^2 & =&
  \frac{e^{-2ik\tau}}{ 2 a^2 }
   k \Big( -1 + i\frac{\gamma(\gamma-1)}{k\tau}
  +  \frac{\gamma(\gamma-1)(\gamma^2-\gamma-1)}{2k^2\tau^2} \nn \\
&& -i  \frac{(\gamma+1)\gamma(\gamma-1)(\gamma-2)(2\gamma^2-2\gamma-3)}{1 2k^3\tau^3}
    +,,,  \Big).
\ea

For the inflation stage,
the  mode solution  of RGW $v_k(\tau)$ of (\ref{u}) at high frequencies
can be expanded as the following
\ba \label{ukexpand}
v_k(\tau)&=&\frac{e^{-ik\tau}}{\sqrt{2k}}
\Big(1-i\frac{\beta(\beta+1)}{2k\tau}
     -\frac{(\beta+2)(\beta+1)\beta(\beta-1)}{8k^2\tau^2}\nn\\
&&+i\frac{(\beta+3)(\beta+2)(\beta+1)\beta(\beta-1)(\beta-2)}{48k^3\tau^3}\nn\\
&&+\frac{(\beta+4)(\beta+3)(\beta+2)(\beta+1)
    \beta(\beta-1)(\beta-2)(\beta-3)}{384k^4\tau^4}      +,,,  \Big),
      \nn \\
\ea
where the  first term is  the   Minkowski spacetime mode [the first term of (\ref{uinfl})],
and the remaining terms reflect the effect of the expanding RW spacetime.
The squared absolute mode is
\ba \label{expusq}
|v_k(\tau)|^2
    &    = & \frac{1}{ 2k}
\Big[ 1 + \frac{\beta(\beta+1)}{2k^2\tau^2}
      +\frac{3(\beta+2)(\beta+1)\beta(\beta-1)}{8 k^4 \tau^4} \nn \\
&&  + \frac{5(\beta+3)(\beta+2)(\beta+1)\beta(\beta-1)(\beta-2)}{16 (k \tau)^6}
            +,,,   \Big] ,
\ea
where the first term $\frac{1}{2k}$ is the Minkowski spacetime vacuum term
giving rise to  $\Delta^2_t(k,\tau) \propto k^2$,
and
the second term $\frac{1}{2k} \frac{\beta(\beta+1)}{2k^2\tau^2}$
gives rise to $\Delta^2_t(k,\tau) \propto   k^0$, respectively.
The squared mode is
\ba \label{expv2}
v_k(\tau)^2
    &    = & \frac{e^{-2ik\tau}}{ 2k}
\Big( 1 -i \frac{\beta(\beta+1)}{k\tau}
      -\frac{(\beta+1)\beta(\beta^2+\beta-1)}{2 k^2 \tau^2} \nn \\
&&  + i\frac{(\beta+2)(\beta+1)\beta(\beta-1)(2\beta^2+2\beta-3)}{12 (k \tau)^3}
   +,,,   \Big)   .
\ea
The time derivatives   are   given by
\ba
( \frac{v_k(\tau)}{a} ) ' & =  & \frac{1}{a}\frac{e^{-ik\tau}}{  \sqrt{2k}} k
\Big( -\frac{i}{k\tau} - \frac{(\beta+1)(\beta+2)}{2k^2\tau^2}
+ i \frac{\beta(\beta+1)(\beta+2)(\beta+3)}{8k^3\tau^3}  \nn  \\
& &   +\frac{(\beta-1)\beta(\beta+1)
   (\beta+2)(\beta+3)(\beta+4)}{48 k^4\tau^4} +,,,   \Big) , \nn
\ea
\ba\label{expdvprim2}
 \Big| ( \frac{v_k(\tau)}{a} ) '  \Big|^2
 &=&
 \frac{k }{2 a^2}
   \Big[1 +\frac{(\beta+1)(\beta+2)}{2k^2\tau^2}
 +\frac{3\beta(\beta+1)(\beta+2)(\beta+3)}{8 k^4\tau^4}  \nonumber  \\
&&  +\frac{5(\beta-1)\beta(\beta+1)(\beta+2)
  (\beta+3)(\beta+4)}{16 k^6 \tau^6}  +,,, \Big],
\ea
\ba \label{vprim}
\Big(( \frac{v_k(\tau)}{a} ) ' \Big)^2
 & =  & \frac{e^{-2ik\tau}}{ 2 a^2}k
\Big( -1 + i \frac{(\beta+1)(\beta+2)}{ k \tau}
  +\frac{(\beta+1)(\beta+2)(\beta^2+ 3\beta+1)}{2 k^2\tau^2} \nn \\
&& -i \frac{\beta (\beta+1)(\beta+2)(\beta+3)(2\beta^2+6\beta+1)}{12 k^3\tau^3}
     +,,,        \Big) . \nn \\
\ea
For de Sitter inflation, the exact expressions are obtained
by letting $\beta+2=0$
in the asymptotic expressions (\ref{ukexpand})--(\ref{vprim}).
We emphasize that the above expressions (\ref{vex4})--(\ref{vprim})
hold only for large $k$.

\section{Continuous Connection of Modes}

Based on the Big-Bang cosmological model,
the durations of four expansion stages after inflation
are specified by the following:
$\frac{a(\tau_H)}{a(\tau_E)}=1.35$,
$\frac{a(\tau_E)}{a(\tau_2)}=2443.0$,
$\frac{a(\tau_2)}{a(\tau_s)}=10^{24}$,
$\frac{a(\tau_s)}{a(\tau_1)}= 20$.
See Ref.\cite{Zhang06A,Zhang06B} for details of the parameters of the scale factor,
where the reheating duration $\frac{a(\tau_s)}{a(\tau_1)}= 300$ was used.
 $a(\tau)$ and $a'(\tau)$ have been chosen to be continuous
at the joining points between adjoining two stages.
From the observed CMB temperature $T_0=2.725$K and
the present Hubble constant
$H_0 \simeq 2.13 \,  h \times 10^{-42}$GeV with $h \simeq 0.7 $,
this set of specification corresponds to
$H\sim 3 \times 10^{14}$GeV for inflation.
Longer durations of reheating and radiation
will lead to higher values of $H$.

The scale factor of inflation is given in (\ref{inflation}).
The mode during  inflation by (\ref{uv}) and (\ref{u}) is
\ba \label{hankelinf}
u_k(\tau ) & = & \sqrt{\frac{\pi}{2}}\sqrt{\frac{x}{2k}}
     \l[ a_1 H^{(1)}_{ \beta+ \frac{1}{2} } (x)
         + a_2 H^{(2)}_{\beta + \frac{1}{2}} ( x) \r],
          \, \,\,\,\,\, \, -\infty  <\tau\leq \tau_1,
\ea
with $ a_1 \equiv A_1  e^{i\pi (\beta +1)/2}$
and $a_2 \equiv  A_2 e^{-i\pi (\beta +1)/2}$.

The scale factor for the reheating  is
\[
a(\tau)= a_z |\tau-\tau_p|^{1+\beta_s} ,\,\,\,\, \tau_1<\tau< \tau_s
\]
with $\beta_s \sim - 0.7$.
The mode during  reheating is written as
\ba  \label{ureh}
u_k(\tau)
& = &
\sqrt{\frac{\pi}{2}}\sqrt{\frac{t}{2k}}
   \l[ b_1   H^{(1)}_{ \beta_s + \frac{1}{2} } (t)
    + b_2     H^{(2)}_{\beta_s + \frac{1}{2}} ( t) \r],
\ea
with $t\equiv k (\tau-\tau_p)$.
By continuous connection of mode and its derivative
at the end of inflation,
 $[u_k(\tau_1)]_{\rm{inf}}=[u_k(\tau_1)]_{\rm{reh}}$
and $[u'_k(\tau_1)]_{\rm{inf}}=[u'_k(\tau_1)]_{\rm{reh}}$,
one gets
{\allowdisplaybreaks
\ba\label{b1ukt3}
b_1&=&
-\frac{\pi }{4 i}\frac{\sqrt{t_1}}{k}\Bigg\{
\sqrt{\frac{x_1}{t_1}}
\left(a_1H^{(1)}_{\beta+\frac{1}{2}}(x_1)
+a_2H^{(2)}_{\beta+\frac{1}{2}}(x_1)\right)
\Big(\frac{k}{2 \sqrt{t_1}} H^{(2)}_{\frac{1}{2}+\beta_s}(t_1)
+\sqrt{t_1} H^{(2)'}_{\frac{1}{2}+\beta_s}(t_1) \Big)
\nn\\
&&
- H^{(2)}_{\frac{1}{2}+\beta_s}(t_1)
  \Big[
 - \frac{k}{2x_1^{1/2}}
\left(a_1H^{(1)}_{\beta+\frac{1}{2}}(x_1)
+a_2H^{(2)}_{\beta+\frac{1}{2}}(x_1)\right)
\nn\\
&&
+ x_1^{1/2}
\left(a_1H^{(1)\,'}_{\beta+\frac{1}{2}}(x_1)
+a_2H^{(2)\,'}_{\beta+\frac{1}{2}}(x_1)\right)
\Big]
\Bigg\} ,
\ea
\ba\label{b2ukt3}
b_2&=&
\frac{\pi }{4 i}\frac{\sqrt{t_1}}{k}
\Bigg\{
\sqrt{\frac{x_1}{t_1}}
\left(a_1H^{(1)}_{\beta+\frac{1}{2}}(x_1)
+a_2H^{(2)}_{\beta+\frac{1}{2}}(x_1)\right)
\Big(\frac{k}{2\sqrt{t_1}} H^{(1)}_{\frac{1}{2}+\beta_s}(t_1)
+\sqrt{t_1}H^{(1)'}_{\frac{1}{2}+\beta_s}(t_1) \Big)
\nn\\
&&
 +H^{(1)}_{\frac{1}{2}+\beta_s}(t_1)
\Big[
\frac{k}{2x_1^{1/2}}
\left(a_1H^{(1)}_{\beta+\frac{1}{2}}(x_1)
+a_2H^{(2)}_{\beta+\frac{1}{2}}(x_1)\right)
\nn\\
&&
-x_1^{1/2}
\left(a_1H^{(1)\,'}_{\beta+\frac{1}{2}}(x_1)
+a_2H^{(2)\,'}_{\beta+\frac{1}{2}}(x_1)\right)
\Big]
\Bigg\} .
\ea
}
In the high frequency limit $k\rightarrow\infty$,
\ba
b_1 &=& e^{i (-x_1-t_1) +i \pi
 (\beta +\beta_s)/2}
 \bigg(
 -1
+i\frac{ \beta  (\beta +1)}{2 x_1}
 +i\frac{ \beta_s (\beta_s+1)}{2 t_1}
+\frac{\beta ^2 (\beta +1)^2}{8 x_1^2}
\nn\\
&&
+\frac{\beta_s^2 (\beta_s+1)^2}{8 t_1^2}
 +\frac{\beta  (\beta +1) \beta_s (\beta_s+1)}{4 t_1 x_1}
\bigg)
+\mathcal O\left(k^{-3}\right)
,
   \\
  \nn \\
b_2 &=& e^{i \left(-x_1+t_1\right)
+i \pi  (\beta -\beta_s)/2}
\left(\frac{\beta  (\beta +1)}{4 x_1^2}
-\frac{\beta_s (\beta_s+1)}{4 t_1^2}\right)
+\mathcal O\left(k^{-3}\right)
.
\ea

During the radiation-dominant stage,
\be
a(\tau)=a_e(\tau-\tau_e),
~~~~~~~
\tau_s\leq\tau\leq \tau_2,
\ee
and the mode function is
\be\label{uradkt}
u_{k}(\tau)=\sqrt{\frac{\pi}{2}}\sqrt{\frac{y}{2k}}
\left[c_1  H^{(1)}_{\frac{1}{2}}(y)
+c_2 H^{(2)}_{\frac{1}{2}}(y)\right],
\ee
with $y  \equiv  k(\tau-\tau_e)$
and the coefficients as
{\allowdisplaybreaks
\ba\label{c1ukt1}
c_1&=&
-\frac{\pi }{4 i}\frac{\sqrt{y_s}}{k}\Bigg\{
\sqrt{\frac{t_s}{y_s}}
\left(b_1H^{(1)}_{\beta_s+\frac{1}{2}}(t_s)
+b_2H^{(2)}_{\beta_s+\frac{1}{2}}(t_s)\right)
 \Big( \frac{k}{2 \sqrt{y_s}} H^{(2)}_{\frac{1}{2}}(y_s)
  +\sqrt{y_s} H^{(2)'}_{\frac{1}{2}}(y_s) \Big)
\nn\\
&&
- H^{(2)}_{\frac{1}{2}}(y_s)
  \Big[
\frac{k}{2t_s^{1/2}}
\left(b_1H^{(1)}_{\beta_s+\frac{1}{2}}(t_s)
+b_2H^{(2)}_{\beta_s+\frac{1}{2}}(t_s)\right)
\nn\\
&&
+t_s^{1/2} \left(b_1H^{(1)\,'}_{\beta_s+\frac{1}{2}}(t_s)
+b_2H^{(2)\,'}_{\beta_s+\frac{1}{2}}(t_s)\right)
\Big]
\Bigg\} ,
\ea
\ba\label{c2ukt1}
c_2&=&
\frac{\pi }{4 i}\frac{\sqrt{y_s}}{k}
\Bigg\{
\sqrt{\frac{t_s}{y_s}}
\left(b_1H^{(1)}_{\beta_s+\frac{1}{2}}(t_s)
+b_2H^{(2)}_{\beta_s+\frac{1}{2}}(t_s)\right)
 \Big(\frac{k}{2\sqrt{y_s}} H^{(1)}_{\frac{1}{2}}(y_s)
+\sqrt{y_s}H^{(1)'}_{\frac{1}{2}}(y_s) \Big)
\nn\\
&&
 - H^{(1)}_{\frac{1}{2}}(y_s)
\Big[ \frac{k}{2t_s^{1/2}}
\left(b_1H^{(1)}_{\beta_s+\frac{1}{2}}(t_s)
+b_2H^{(2)}_{\beta_s+\frac{1}{2}}(t_s)\right)
\nn\\
&&
+ t_s^{1/2}
\left(b_1H^{(1)\,'}_{\beta_s+\frac{1}{2}}(t_s)
+b_2H^{(2)\,'}_{\beta_s+\frac{1}{2}}(t_s)\right)
\Big]
\Bigg\} ,
\ea
}
where $t_s\equiv k(\tau_s-\tau_p)$ and $y_s\equiv k(\tau_s-\tau_e)$.
In the limit $k\rightarrow\infty$
{\allowdisplaybreaks
\ba
c_1  &=&
e^{
i (-x_1-t_1+t_s-y_s)
+i \pi  \beta/2}
\bigg(
-1
+i\frac{ \beta  (\beta +1)}{2 x_1}
+i\frac{ \beta_s (\beta_s+1)}{2 t_1}
-i\frac{ \beta_s (\beta_s+1)}{2 t_s}
\nn\\
&&
+\frac{\beta ^2 (\beta +1)^2}{8 x_1^2}
+\frac{\beta_s^2 (\beta_s+1)^2}{8 t_1^2}
+\frac{\beta_s^2 (\beta_s+1)^2}{8 t_s^2}
+\frac{\beta  (\beta +1) \beta_s (\beta_s+1)}{4 t_1 x_1}
\nn\\
&&
-\frac{\beta  (\beta +1) \beta_s (\beta_s+1)}{4 x_1 t_s}
-\frac{\beta_s^2 (\beta_s+1)^2}{4 t_1 t_s}
\bigg)
+\mathcal O\left(k^{-3}\right) ,
    \\
    \nn\\
c_2 &=&
\left(\frac{\beta  (\beta +1)}{4 x_1^2}
-\frac{\beta_s (\beta_s+1)}{4 t_1^2}\right)
e^{i (-x_1+t_1-t_s+y_s)+i \pi  \beta /2}
\nn\\
&&
+\frac{\beta_s (\beta_s+1) }{4 t_s^2}
e^{
i (-x_1-t_1+t_s+y_s)+i \pi  \beta /2}
+O\left(k^{-3}\right) .
\ea
}

During the matter-dominant stage,
\be
a(\tau)=a_m(\tau-\tau_m)^{2},
~~~~~~~
\tau_2\leq\tau\leq \tau_E,
\ee
and
the mode function is
\be\label{umatkt}
u_{k}(\tau)=\sqrt{\frac{\pi}{2}}\sqrt{\frac{z}{2k}}
\left[d_1 H^{(1)}_{\frac{3}{2}}(z)
    +d_2 H^{(2)}_{\frac{3}{2}}(z)\right],
\ee
with $z \equiv k(\tau-\tau_m)$.
The coefficients are
{\allowdisplaybreaks
\ba\label{d1ukt1}
d_1&=&
-\frac{\pi }{4 i}\frac{\sqrt{z_2}}{k}\Bigg\{
\sqrt{\frac{y_2}{z_2}}
\left(c_1H^{(1)}_{\frac{1}{2}}(y_2)
+c_2H^{(2)}_{\frac{1}{2}}(y_2)\right)
  \Big(\frac{k}{2 \sqrt{z_2}} H^{(2)}_{\frac{3}{2}}(z_2)
+\sqrt{z_2} H^{(2)'}_{\frac{3}{2}}(z_2) \Big)
          \nn\\
&&
- H^{(2)}_{\frac{3}{2}}(z_2)
  \Big[
\frac{k}{2y_2^{1/2}}
\left(c_1H^{(1)}_{\frac{1}{2}}(y_2)
+c_2H^{(2)}_{\frac{1}{2}}(y_2)\right)
\nn\\
&&
+y_2^{1/2} \left(c_1H^{(1)\,'}_{\frac{1}{2}}(y_2)
+c_2H^{(2)\,'}_{\frac{1}{2}}(y_2)\right)
\Big]
\Bigg\} ,
\ea
\ba\label{d2ukt1}
d_2&=&
\frac{\pi }{4 i}\frac{\sqrt{z_2}}{k}
\Bigg\{
\sqrt{\frac{y_2}{z_2}}
\left(c_1H^{(1)}_{\frac{1}{2}}(y_2)
+c_2H^{(2)}_{\frac{1}{2}}(y_2)\right)
 \Big( \frac{k}{2\sqrt{z_2}} H^{(1)}_{\frac{3}{2}}(z_2)
   +\sqrt{z_2}H^{(1)'}_{\frac{3}{2}}(z_2) \Big)
\nn\\
&&
 - H^{(1)}_{\frac{3}{2}}(z_2)
\Big[ \frac{k}{2y_2^{1/2}}
\left(c_1H^{(1)}_{\frac{1}{2}}(y_2)
+c_2H^{(2)}_{\frac{1}{2}}(y_2)\right)
  \nn\\
&&
+ y_2^{1/2}
\left(c_1H^{(1)\,'}_{\frac{1}{2}}(y_2)
+c_2H^{(2)\,'}_{\frac{1}{2}}(y_2)\right)
\Big]
\Bigg\} ,
\ea
}
where $y_2\equiv k(\tau_2-\tau_e)$ and $z_2\equiv k(\tau_2-\tau_m)$.
In the limit $k\rightarrow\infty$,
{\allowdisplaybreaks
\ba
d_1 &=&
i \bigg(
-1
+i\frac{ \beta  (\beta +1)}{2 x_1}
+i\frac{ \beta_s (\beta_s+1)}{2 t_1}
-i\frac{ \beta_s (\beta_s+1)}{2 t_s}
+\frac{i}{z_2}
+\frac{\beta ^2 (\beta +1)^2}{8 x_1^2}
\nn\\
&&
+\frac{\beta_s^2 (\beta_s+1)^2}{8 t_1^2}
+\frac{\beta_s^2 (\beta_s+1)^2}{8 t_s^2}
+\frac{1}{2 z_2^2}
+\frac{\beta  (\beta +1) \beta_s (\beta_s+1)}{4 x_1 t_1}
\nn\\
&&
-\frac{\beta  (\beta +1) \beta_s (\beta_s+1)}{4 x_1 t_s}
+\frac{\beta  (\beta +1)}{2 x_1 z_2}
-\frac{\beta_s^2 (\beta_s+1)^2}{4 t_1 t_s}
+\frac{\beta_s (\beta_s+1)}{2 t_1 z_2}
\nn\\
&&
-\frac{\beta_s (\beta_s+1)}{2 t_s z_2}
\bigg)
e^{i (-x_1-t_1+t_s-y_s+y_2-z_2)+{i \pi  \beta }/{2}}
+\mathcal O\left(k^{-3}\right)
,
    \\
    \nn\\
d_2 &=&
i \left(-\frac{\beta  (\beta +1)}{4 x_1^2}
+\frac{\beta_s (\beta_s+1)}{4 t_1^2}
\right)
e^{i (-x_1 +t_1-t_s+y_s-y_2+z_2)+{i \pi  \beta }/{2}}
\nn\\
&&
-i \frac{\beta_s (\beta_s+1)}{4 t_s^2}
e^{i (-x_1-t_1+t_s+y_s-y_2+z_2)+{i \pi  \beta }/{2}}
\nn\\
&&
+\frac{i}{2 z_2^2}
e^{i (-x_1-t_1+t_s-y_s+y_2+z_2)+{i \pi  \beta }/{2}}
+\mathcal O\left(k^{-3}\right)
.
\ea
}

The present accelerating  stage has
\be
a(\tau)=l_H|\tau-\tau_a|^{-\gamma},
~~~~~~~
\tau_E\leq\tau\leq \tau_H,
\ee
with $\gamma\simeq 2.108$,
fits the model $\Omega_\Lambda \simeq 0.7$ and $\Omega_m =1-\Omega_\Lambda $.
The normalization of $a(\tau)$ is taken such that $|\tau_H-\tau_a|=1$
and $a(\tau_H)=l_H$.
The present Hubble constant is $H_0= (a'/a^2)|_{\tau_H} =
\gamma\,  l_H^{-1}$.
The mode function is
\be\label{umatkt2}
u_{k}(\tau)=\sqrt{\frac{\pi}{2}}\sqrt{\frac{s}{2k}}
\left[e^{-i\pi\gamma/2}\beta_k H^{(1)}_{-\gamma-\frac{1}{2}}(s)
+e^{i\pi\gamma/2}\alpha_k H^{(2)}_{-\gamma-\frac{1}{2}}(s)\right],
\ee
with $s \equiv k|\tau-\tau_a|=-k(\tau-\tau_a)$
and
{\allowdisplaybreaks
\ba\label{betakukt1}
e^{-i\pi\gamma/2}\beta_k &=&
\frac{\pi }{4 i}\frac{\sqrt{s_E}}{k}
\Bigg\{
\sqrt{\frac{z_E}{s_E}}
\left(d_1H^{(1)}_{\frac{3}{2}}(z_E)
+d_2H^{(2)}_{\frac{3}{2}}(z_E)\right)
  \Big( -\frac{k}{2 z_E^{1/2}}
H^{(2)}_{-\gamma-\frac{1}{2}}(s_E)
+z_E^{1/2} H^{(2)'}_{-\gamma-\frac{1}{2}}(s_E)
    \Big)  \nn \\
&& - H^{(2)}_{-\gamma-\frac{1}{2}}(s_E)
  \Big[
\frac{k}{2z_E^{1/2}}
\left(d_1H^{(1)}_{\frac{3}{2}}(z_E)
+d_2H^{(2)}_{\frac{3}{2}}(z_E)\right)
\nn\\
&&
+z_E^{1/2} \left(d_1H^{(1)\,'}_{\frac{3}{2}}(z_E)
+d_2H^{(2)\,'}_{\frac{3}{2}}(z_E)\right)
\Big]
\Bigg\} ,
\ea
\ba\label{alphakukt1}
e^{i\pi\gamma/2}\alpha_k&=&
\frac{\pi }{4 i}\frac{\sqrt{s_E}}{k}
\Bigg\{
\sqrt{\frac{z_E}{s_E}}
\left(d_1H^{(1)}_{\frac{3}{2}}(z_E)
+d_2H^{(2)}_{\frac{3}{2}}(z_E)\right)  \Big(
\frac{k}{2 z_E^{1/2} }
H^{(1)}_{-\gamma-\frac{1}{2}}(s_E)
-z_E^{1/2} H^{(1)'}_{-\gamma-\frac{1}{2}}(s_E)
\Big)   \nn \\
&&    +H^{(1)}_{-\gamma-\frac{1}{2}}(s_E)
\Big[
\frac{k}{2z_E^{1/2}}
\left(d_1H^{(1)}_{\frac{3}{2}}(z_E)
+d_2H^{(2)}_{\frac{3}{2}}(z_E)\right)
        \nn\\
&& + z_E^{1/2}
\left(d_1H^{(1)\,'}_{\frac{3}{2}}(z_E)
+d_2H^{(2)\,'}_{\frac{3}{2}}(z_E)\right)
\Big]
\Bigg\} ,
\ea
}
where $z_E\equiv k(\tau_E-\tau_m)$ and $s_E\equiv -k(\tau_E-\tau_a)$.
In the  limit $k\rightarrow\infty$,
the expansion of $\beta_k$ and $\alpha_k$ are given by (\ref{e1})
(\ref{e2}).

\end{document}